\documentclass[letterpaper,journal,twocolumn]{IEEEtran}
\usepackage{cite}
\usepackage{amsmath,amssymb,amsthm,amsfonts}
\usepackage{algorithm,algpseudocode}
\usepackage{graphicx}
\usepackage{booktabs,threeparttable}
\usepackage{multirow}
\usepackage{textcomp}
\usepackage{xcolor}
\usepackage{stfloats}
\usepackage[nolist]{acronym}
\usepackage{bm}
\usepackage{relsize}
\usepackage{hyperref}
\hypersetup{hidelinks,
	colorlinks=true,
	allcolors=black,
	pdfstartview=Fit,
	breaklinks=true
}

\newcommand{\HD}{{\rm HD}}
\newcommand{\PM}{{\rm PM}}

\newcommand{\bigboxplus}{\mathlarger{\boxplus}}
\newcommand{\Idx}[1]{[\![#1]\!]}
\newcommand{\s}[1]{\!#1\!}
\newcommand{\Rmnum}[1]{\uppercase\expandafter{\romannumeral #1}}

\newtheorem{remark}{Remark}

\begin{document}

\title{Node-Based Soft-Output Fast Successive Cancellation List Decoding of Polar Codes}

\author{
	{Li Shen}, \IEEEmembership{Graduate Student Member, IEEE},
	{Yongpeng Wu}, \IEEEmembership{Senior Member, IEEE},
	{Zhen Gao}, \IEEEmembership{Senior Member, IEEE},
	{Yin Xu}, \IEEEmembership{Senior Member, IEEE},
	{Xiaohu You}, \IEEEmembership{Fellow, IEEE},
	{Xiqi Gao}, \IEEEmembership{Fellow, IEEE},
	and {Wenjun Zhang},~\IEEEmembership{Fellow, IEEE}

\thanks{This paper has been presented in part at the IEEE Wireless Communications and Networking Conference (WCNC) 2025 \cite{Shen2025Soft}. The work of Yongpeng Wu was supported in part by the Fundamental Research Funds for the Central Universities, the Yangtze River Delta Science and Technology Innovation Community Joint Research (Basic Research) Project under Grant BK20244006, 111 project BP0719010, and STCSM 22DZ2229005. The work of Zhen Gao was supported in part by the Natural Science Foundation of China (NSFC) under Grant 62471036 and Grant U2233216, in part by Shandong Province Natural Science Foundation under Grant ZR2025QA30, in part by Beijing Natural Science Foundation under Grants L242011, QY24167, QY25256, QY25257. The work of Yin Xu was funded in part by the National Natural Science Foundation of China under Grant 62422111, Grant 62371291 and Grant 62431015. The work of  Xiqi Gao was supported by the Jiangsu Province Major Science and Technology Project under Grant BG2024005. \emph{(Corresponding author: Yongpeng Wu.)}}
\thanks{Li Shen, Yongpeng Wu, Yin Xu, and Wenjun Zhang are with the School of Integrated Circuits (School of Information Science and Electronic Engineering), Shanghai Jiao Tong University, Shanghai 200240, China (e-mail: \href{mailto:shen-l@sjtu.edu.cn}{shen-l@sjtu.edu.cn}; \href{mailto:yongpeng.wu@sjtu.edu.cn}{\mbox{yongpeng.wu@sjtu.edu.cn}}; \href{mailto:xuyin@sjtu.edu.cn}{xuyin@sjtu.edu.cn}; \href{mailto:zhangwenjun@sjtu.edu.cn}{zhangwenjun@sjtu.edu.cn}).}
\thanks{Zhen Gao is with  Beijing Institute of Technology (BIT), Zhuhai 519088, China, also with the State Key Laboratory of CNS/ATM, Beijing 100081, China, also with the MIIT Key Laboratory of Complex-Field Intelligent Sensing, Beijing 100081, China, also with the Advanced Technology Research Institute, BIT, Jinan 250307, China, and also with the Yangtze Delta Region Academy, BIT, Jiaxing 314019, China (e-mail: \href{gaozhen16@bit.edu.cn}{gaozhen16@bit.edu.cn}).}
\thanks{Xiaohu You and Xiqi Gao are with the National Mobile Communications Research Laboratory, Southeast University, Nanjing 210096, China and are also with Purple Mountain Laboratories, Nanjing 211111, China (e-mail: \href{mailto:xhyu@seu.edu.cn}{xhyu@seu.edu.cn}; \href{mailto:xqgao@seu.edu.cn}{xqgao@seu.edu.cn})}
}

\begin{acronym}
	\acro{AWGN}{additive white Gaussian noise}
	\acro{APP}{a-posteriori probability}
	\acro{BCJR}{Bahl-Cocke-Jelinek-Raviv}
	\acro{BER}{bit error rate}
	\acro{BLER}{block error rate}
	\acro{BP}{belief propagation}
	\acro{CRC}{cyclic redundancy check}
	\acro{eBCH}{extended Bose-Chaudhuri-Hocquenghem}
	\acro{FastSOL}{fast soft-output list}
	\acro{FIC}{frozen-information cascade}
	\acro{FSCL}{fast successive cancellation list}
	\acro{GLDPC}{generalized low-density parity-check}
	\acro{G-REP}{generalized REP}
	\acro{G-PC}{generalized parity-check}
	\acro{HWF}{hardware-friendly}
	\acro{LLR}{log-likelihood ratio}
	\acro{MAP}{maximum a posteriori}
	\acro{MIMO}{multiple-input multiple-output}
	\acro{ML}{maximum-likelihood}
	\acro{MMSE}{minimum mean-square error}
	\acro{PM}{path metric}
	\acro{QPSK}{quadrature phase shift keying}
	\acro{Rate0}{rate-zero}
	\acro{Rate1}{rate-one}
	\acro{REP}{repetition}
	\acro{RV}{random variable}
	\acro{SC}{successive cancellation}
	\acro{SCAN}{soft cancellation}
	\acro{SCL}{successive cancellation list}
	\acro{SO}{soft output}
	\acro{SOGRAND}{soft-output guessing random additive noise decoding}
	\acro{SO-SCL}{soft-output successive cancellation list}
	\acro{SO-FSCL}{soft-output fast successive cancellation list}
	\acro{SPC}{single-parity-check}
	\acro{SR0/REP}{sequence Rate0 or REP}
	\acro{SR1/SPC}{sequence Rate1 or SPC}
\end{acronym}

\maketitle

\begin{abstract}
The soft-output successive cancellation list (SO-SCL) decoder provides a methodology for estimating the a-posteriori probability log-likelihood ratios by only leveraging the conventional SCL decoder of polar codes. However, the sequential decoding nature of SCL introduces high decoding latency to SO-SCL. In this paper, we incorporate node-based fast decoding into the SO-SCL framework. After addressing the challenge of soft output extraction in special node decoding, we proposed the soft-output fast SCL (SO-FSCL) decoding algorithm, along with its log-domain implementation and hardware-friendly version. The proposed SO-FSCL decoder can be regarded as an add-on extension to FSCL decoder, enabling us to autonomously choose whether to output only hard decisions like FSCL or to provide additional soft outputs. Latency and complexity analyses demonstrate that SO-FSCL can significantly reduce, for example, decoding time steps by 81.8\% (with unlimited resources), the number of additions by 41.3\%, and the number of comparisons by 46.4\%. Meanwhile, simulation results indicate that SO-FSCL delivers almost the same soft-output performance as SO-SCL, outperforming other soft-output polar decoders, especially in scenarios involving iterative decoding.
\end{abstract}

\section{Introduction}
Polar codes are a state-of-the-art channel coding scheme that leverage the concept of channel polarization \cite{Arikan2009Channel}. The capacity-achieving capability of polar codes is accomplished by performing the \ac{SC} decoding algorithm \cite{Arikan2009Channel}. However, insufficient polarization at a practical moderate-to-short block length will broaden the performance gap between \ac{SC} decoder and \ac{ML} decoder \cite{Tal2015List}. To alleviate this issue, the \ac{SCL} decoder \cite{Tal2015List} outputs a list of the most probable candidate codewords for selection. By concatenating \ac{CRC} codes as the outer codes and detecting which candidate codeword is correct, the \ac{CRC}-aided \ac{SCL} decoding \cite{Tal2015List,Niu2012CRC,Balats2015LLR} can lead to a dramatic performance enhancement over conventional \ac{SCL} decoding at moderate-to-short block lengths. Owing to their superiority, 3GPP has incorporated polar codes as the channel coding solution for control channels in 5G communication systems \cite{3GPP212}.

In the future 6G communication systems, the support for diverse novel features will impose more stringent requirements on technical indicators like end-to-end latency, throughput, and reliability \cite{Rowshan2024Channel,wu2024physical}. However, the sequential bit-by-bit decoding structure of conventional \ac{SC} and \ac{SCL} decoders obstruct the implementation of lower-latency polar decoding. Hence, many researches have been devoted to fast \ac{SC}/\ac{SCL} decoding via identifying some special subcodes (nodes) in the recursive manner of polar codes \cite{Alamdar2011Simp,Sarkis2014Fast,Hanif2017Fast,Condo2018Generalized,Zheng2021Threshold,Lu2024Fast,Sarkis2016Fast,Hashemi2017Fast,Ardakani2019Fast,Ren2022Sequence,Lu2025Fast}, where the codewords of special subcodes are directly estimated by exploring the characteristics of these subcodes. 

Specifically, \cite{Alamdar2011Simp} first proposed a simplified \ac{SC} decoder by processing \ac{Rate0} and \ac{Rate1} nodes, which contain no information bits and no frozen bits, respectively. In \cite{Sarkis2014Fast}, the authors further identify \ac{SPC} nodes, \ac{REP} nodes, and some of their combinations, e.g., \ac{REP}-\ac{SPC}, \ac{Rate0}-\ac{SPC}, and \ac{Rate0}-\ac{Rate1} nodes. The four basic node types mentioned above were subsequently incorporated as special cases of the \mbox{Type-\Rmnum{1}} to Type-\Rmnum{5} nodes proposed in \cite{Hanif2017Fast}. More general types of special nodes have been explored in \cite{Condo2018Generalized,Zheng2021Threshold,Lu2024Fast} for \ac{SC} decoder, such as \ac{G-REP} nodes, \ac{G-PC} nodes, \ac{SR0/REP} nodes, extended \ac{G-PC} nodes, and \ac{SR1/SPC} nodes. Moreover, the idea of node-based fast decoding has naturally been extended to \ac{FSCL} decoder in \cite{Condo2018Generalized,Sarkis2016Fast,Hashemi2017Fast,Ardakani2019Fast,Ren2022Sequence,Lu2025Fast}, by addressing the path splitting and selection processes at the codeword side of these considered nodes to enable list decoding. Benefiting from the introduction of special nodes, these fast decoding algorithms significantly enhance the parallelism of \ac{SC}/\ac{SCL} decoding while typically incurring no or negligible performance degradation.

However, the aforementioned polar decoding algorithms are actually hard-output and unable to yield the post-decoding bit-wise \acp{SO}, which hinders the utilization of polar codes in the field of iterative detection and iterative decoding. For example, although \cite{dai2018polar} proposed a joint polar coding and \ac{MIMO} transmission framework that outperforms mainstream coded \ac{MIMO} schemes, it fails to support \ac{MIMO} iterative detection and decoding \cite{Hochwald2003Ach} due to its hard-output nature. Furthermore, \acp{SO} are crucial for utilizing polar codes in more iterative scenarios, including serving as component codes in \ac{GLDPC} codes \cite{Liva2008Quasi, shen2025GLDPC} or product codes \cite{Elias1954Error}, as well as enhancing code-aided \ac{MIMO} channel estimation \cite{Shi2024Code}.

Although the \ac{BCJR} algorithm \cite{Bahl1974Optimal} can output an optimal estimate of the \ac{APP} that a decoded bit is correct or not, its complexity is exponential and thus it is impractical. Along the factor graph of polar codes, polar \ac{BP} decoder \cite{Arikan2008Perform} and \ac{SCAN} decoder \cite{Fayyaz2014Low} could offer \acp{SO}. However, both of them and some algorithms based on them, e.g. \cite{Yuan2014Early,Elkelesh2018Belief,Pillet2020SCAN}, require inner iterations. Even so, the short girths in the factor graph still deteriorate the accuracy of the estimated \acp{APP} and their error correction performance. To achieve a decoding performance similar to \ac{SCL} decoding while providing \acp{SO}, \cite{Xiang2020Soft,Egilmez2022soft,Fominykh2023Effic} proposed to cascade a \ac{BP}, \ac{SCAN}, or sum-product decoder after a \ac{SCL} decoder, respectively. Through the method of sparsifying Tanner graph to eliminate some short cycles in \cite{Li2024Belief}, \ac{BP} decoding can also achieve comparable performance to \ac{SCL} for certain short polar codes. Moreover, node-based fast decoding for \ac{SCAN} decoder has been considered in \cite{Pillet2021Fast}. By identifying \ac{Rate0}, \ac{Rate1} and \ac{REP} nodes, \cite{Shen2022Fast} proposed a \ac{FastSOL} decoder that integrates \ac{SCAN} and \ac{SCL} into a single decoding procedure, at the cost of a little error correction performance and complexity.

Disregarding \ac{BP} and \ac{SCAN} decoders, Pyndiah proposed estimating bit-wise \acp{APP} from a list of candidate codewords \cite{Pyndiah1998Near}, and the \ac{SCL} decoding can exactly provide such a list. Thus, this approximation has been applied for iterative decoding of polar product codes in \cite{Condo2020Prac, Coskun2024Prec}. For any codes with moderate-to-short redundancy, a generalized \ac{SOGRAND} was presented \cite{Galligan2023Upgrade,Yuan2025Soft}, which avoids the shortcoming that the previous Pyndiah's approximation \cite{Pyndiah1998Near} ignores the \acp{APP} of codewords not included in the list. Then, this idea has been extended to guessing codeword decoding \cite{Duffy2025Soft} and \ac{SCL} decoding \cite{Yuan2024Near,Yuan2025SoSCL}. Leveraging the \ac{SCL} decoding tree, the \ac{SO-SCL} decoder \cite{Yuan2025SoSCL} has demonstrated a bit-wise \ac{SO} performance close to the \ac{BCJR} algorithm and a complexity comparable to \ac{SCL} decoding.

In this paper, we investigate how to further accelerate \ac{SO-SCL} decoding with the assistance of node-based fast decoding and proposed a \ac{SO-FSCL} decoder. Since the extraction of \acp{SO} in \ac{SO-SCL} requires accessing all roots of unvisited subtrees in the \ac{SCL} decoding tree \cite{Yuan2025SoSCL}, while the node-based fast decoding may only visit some of them \cite{Hashemi2017Fast,Ardakani2019Fast,Ren2022Sequence}, we need to address this primary challenge for \ac{SO-FSCL}. Besides, compatibility with dynamic frozen bits to enhance \ac{SO} accuracy \cite{Yuan2025SoSCL} is also worth considering. The main contributions are summarized as follows:
\begin{itemize}
    \item To facilitate the extraction of \acp{SO} and ensure compatibility with dynamic frozen bits, we constrained the generalized special nodes to the form of \ac{FIC}. Subsequently, a \ac{SO} node decoding method for \ac{FIC} nodes is proposed, which can be regarded as an add-on extension of hard-output \ac{FSCL} decoding.
    \item By incorporating compatibility with dynamic frozen bits and modification for bit-wise \ac{APP} \acp{LLR}, we proposed the \ac{SO-FSCL} decoding algorithm. Furthermore, a log-domain implementation and a \ac{HWF} version of SO-FSCL decoding are presented.
    \item Analysis of the decoding latency (under both unlimited and limited resource constraints) and complexity performance of \ac{SO-FSCL} is conducted, demonstrating its significant advantages over \ac{SO-SCL} in both aspects.
    \item Simulation results show that the \ac{SO} performance of \ac{SO-FSCL} is identical to that of \ac{SO-SCL}. Moreover, we highlight the applications of \ac{SO-FSCL} in \ac{MIMO} iterative systems and \ac{GLDPC} decoders.
\end{itemize}

Note that a preliminary version of this work has been presented in \cite{Shen2025Soft}. Compared to the conference version, the supplementary contributions of this paper are summarized as follows. First, we extend the \ac{SO-FSCL} decoding scheme, originally based on four specific node types, to more generalized nodes, and propose corresponding \ac{SO} decoding method for the considered nodes. Second, we propose a modification to the calculation of the output \ac{APP} \acp{LLR}. Moreover, considering practical implementation, we also present the log-domain and \ac{HWF} versions of {SO-FSCL} decoding. Finally, a more comprehensive analysis is conducted on the decoding latency, complexity, and performance.

The rest of the paper is organized as follows. Section \Rmnum{2} provides some essential preliminaries. In Section \Rmnum{3}, we introduce the proposed \ac{SO-FSCL} decoder. Then, we analyze the decoding latency and complexity performance in Section \Rmnum{4}. Section \Rmnum{5} presents simulation results and applications of \ac{SO-FSCL} decoding. Finally, Section \Rmnum{6} concludes this paper.

\section{Preliminaries}
\subsection{Notations}
In this paper, \acp{RV} are denoted by uppercase letters, e.g., $X$, and their realizations are denoted by corresponding lowercase letters, e.g., $x$. We utilize bold letters to indicate vectors or matrices, whereas sets are distinguished by calligraphic fonts. A vector of length $N$ is denoted as $\bm{x}^N = (x[1], x[2], \cdots, x[N])$, where $x[i]$ is the $i$-th entry. We write $\bm{x}[i\s{:}j]$ to denote the subvector $(x[i], x[i+1], \cdots, x[j])$, $i \leq j$. Given an index set $\mathcal{A} \subseteq \{1,2,\cdots,N\}$, $\bm{x}[\mathcal{A}]$ denotes the subvector consisting of all $x[i]$, $i \in \mathcal{A}$. The notations $\mathcal{X}^C$ and $|\mathcal{X}|$ represent the complement and cardinality of a set $\mathcal{X}$, respectively. The probability density function of a continuous \ac{RV} $X$ and the probability mass function of a discrete \ac{RV} $Y$ are denoted as $p_X$ and $P_Y$, respectively. An index set $\{i, i\s{+}1, \cdots,j\}$ is abbreviated as $\Idx{i\s{:}j}$, $i \leq j$, and $\Idx{1\s{:}j}$ is further abbreviated as $\Idx{j}$. For completeness, we stipulate that $\bm{x}[i\s{:}j]$ is void and $\Idx{i\s{:}j} = \emptyset$ when $i > j$.

\subsection{Polar Codes}
We define that an $(N, K)$ polar code is of code length $N=2^n$ and code dimension $K$, where $n$ is a positive integer. According to the channel polarization theory \cite{Arikan2009Channel}, the $K$ most reliable polarized subchannels out of $N$ are used for message transmission, with their indices denoted as $\mathcal{I} \subseteq \Idx{N}$. The remaining subchannels are indexed by $\mathcal{F} = \Idx{N} \cap \mathcal{I}^C$. Thus, the input vector $\bm{u}^N$ for polar transform consists of $\bm{u}[\mathcal{I}]$ and $\bm{u}[\mathcal{F}]$, which are placed with information bits and frozen bits, respectively. Each frozen bit $u[i]$, $i \in \mathcal{F}$, is either set to a static value, like zero, or determined as a linear function of previous input $\bm{u}[1\s{:}i-1]$, which is also known as the \textit{dynamic frozen bit} \cite{Trifonov2016Polar}. Then, the polarization or encoding process is mathematically expressed as
\begin{equation}
    \bm{c}^N = \bm{u}^N \bm{G}_N,
\end{equation}
where the generator matrix $\bm{G}_N = \bm{G}_2^{\otimes n}$ is the $n$-th Kronecker power of Arıkan's kernel $\bm{G}_2 = \left[\begin{smallmatrix}1&0\\1&1\end{smallmatrix}\right]$.

\subsection{SC, SCL and Their Fast Decoding}
\begin{figure}[!t]
    \centering
    \includegraphics[width=0.48\textwidth]{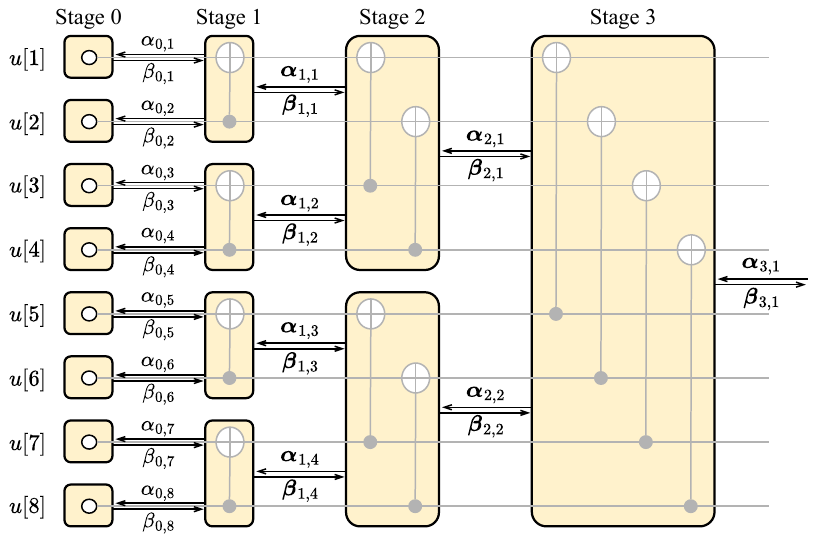}
    \caption{Factor graph of polar codes with $N=8$.}
    \label{fig:PolarTree}
    \vspace{-10pt}
\end{figure}

At the receiver, the \ac{SC} decoder estimates each bit in $\bm{u}^N$ sequentially, along the polar factor graph as shown in Fig.~\ref{fig:PolarTree}. The factor graph has $n$ stages, where the $\psi$-th stage contains $2^{n-\psi}$ nodes, marked with yellow rectangles in Fig.~\ref{fig:PolarTree}. Thus, these nodes constitute a tree rooted at the $n$-th stage. By the recursive property of polar codes \cite{Arikan2009Channel}, each node at the $\psi$-th stage could be interpreted as a sub-polar code of length $2^\psi$. For the $\phi$-th node at the $\psi$-th stage, it receives the internal \ac{LLR} vector $\bm{\alpha}_{\psi,\phi}$ from its parent node and feeds back the internal bit vector $\bm{\beta}_{\psi,\phi}$. Specifically, the internal \acp{LLR} passed from a parent node to its two child nodes are computed as \cite{Balats2015LLR}
\begin{align}
    &\alpha_{\psi\s{-}1,2\phi\s{-}1}[j] = f(\alpha_{\psi,\phi}[j], \alpha_{\psi,\phi}[j\s{+}2^{\psi\s{-}1}]), \label{eq:f_func}\\
    &\alpha_{\psi\s{-}1,2\phi}[j] = \left(1\s{-}2\beta_{\psi\s{-}1,2\phi\s{-}1}[j]\right)\alpha_{\psi,\phi}[j] \s{+} \alpha_{\psi,\phi}[j\s{+}2^{\psi\s{-}1}], \label{eq:g_func}
\end{align}
where $j \in \Idx{2^{\psi-1}}$, and the $f$-function is defined as
\begin{equation}
\begin{aligned}
    f(a,b) &\triangleq \ln\left(\frac{e^{a+b}+1}{e^a+e^b}\right) \\
    &= {\rm sign}(a) {\rm sign}(b) \min\{|a|,|b|\} + \\ 
    &\quad\; \ln\left(1+e^{-|a+b|}\right) - \ln\left(1+e^{-|a-b|}\right).
\end{aligned}
\end{equation}
On the other hand, the internal bit vector $\bm{\beta}_{\psi,\phi}$ is updated by the two child nodes as \cite{Balats2015LLR}
\begin{align}
    &\beta_{\psi,\phi}[j] = \beta_{\psi\s{-}1,2\phi\s{-}1}[j] \oplus \beta_{\psi\s{-}1,2\phi}[j], \label{eq:b1_func} \\
    &\beta_{\psi,\phi}[j\s{+}2^{\psi\s{-}1}] = \beta_{\psi\s{-}1,2\phi}[j]. \label{eq:b2_func}
\end{align}
In particular, the \ac{LLR} vector $\bm{\alpha}_{n,1}$ at the $n$-th stage is initialized as the channel \acp{LLR} 
\begin{equation}
    \ell_{\rm ch}[i] \triangleq \ln\frac{p_{Y|C}(y[i]|0)}{p_{Y|C}(y[i]|1)}, i\in\Idx{N},
\end{equation}
given the channel observation $\bm{y}^N$. At the $0$-th stage, the $i$-th estimated bit $\hat{u}[i]$ is determined by
\begin{equation}
    \hat{u}[i] = \beta_{0,i} = \left\{\begin{aligned}
        &\text{frozen value}, && i\in\mathcal{F}; \\
        &\HD(\alpha_{0,i}), && i\in\mathcal{I},
    \end{aligned}\right.
\end{equation}
where $\HD(x)$ denotes the hard decision function that outputs 0 if $x \geq 0$ and 1 otherwise. 

The \ac{SCL} decoder diverges from the \ac{SC} decoder by evaluating both binary hypotheses (0 and 1) for each information bit, instead of exclusively maintaining the most probable bit estimation. Specifically, given a list size $L$, the candidate estimates $\hat{\bm{u}}^N$ (decoding paths) double at each decision on $\hat{u}[i]$ for $i \in \mathcal{I}$, and only $L$ paths with the lowest \acp{PM} survive. After the $i$-th bit decision, the \ac{PM} associated with the $l$-th path, denoted by $\PM_i^{(l)}$, is calculated as \cite{Balats2015LLR}
\begin{equation}
\begin{aligned}
    \PM_i^{(l)} &= \sum_{k=1}^{i} \ln\left( 1 + e^{-\left(1-2\hat{u}_l[k]\right)\alpha_{0,k}^{(l)}} \right)\\
    &= \PM_{i-1}^{(l)} + \ln\left( 1 + e^{-\left(1-2\hat{u}_l[i]\right)\alpha_{0,i}^{(l)}} \right),
\end{aligned}
\label{eq:PM_u}
\end{equation}
where $\hat{\bm{u}}_l^N$ is the estimated input vector at the $l$-th path, and $\alpha_{0,k}^{(l)}$ is the internal \ac{LLR} $\alpha_{0,k}$ corresponding to this path. Note that although there is no path splitting when decoding a frozen bit, the \acp{PM} still need to be updated accordingly.

To reduce the latency of \ac{SC} and \ac{SCL} decoding, the studies in \cite{Alamdar2011Simp,Sarkis2014Fast,Hanif2017Fast,Condo2018Generalized,Zheng2021Threshold,Lu2024Fast,Sarkis2016Fast,Hashemi2017Fast,Ardakani2019Fast,Ren2022Sequence,Lu2025Fast} have proposed the node-based fast decoding algorithms by identifying special nodes in the factor graph. Considering the $\phi_s$-th node at the $\psi_s$-th stage, the corresponding indices in $\bm{u}^N$ at $0$-th stage underneath this node start at $i_s = (\phi_s\s{-}1)2^{\psi_s}+1$ and end at $j_s = \phi_s2^{\psi_s}$. Such a node is denoted by $\mathbb{N}_{i_s}^{j_s}$ hereafter. Since $\mathbb{N}_{i_s}^{j_s}$ could be regarded as a sub-polar code of length $N_s = j_s \s{-} i_s \s{+} 1$, the sub-codeword is then generated by $\bm{s}^{N_s} = \bm{u}[i_s\s{:}j_s] \bm{G}_{N_s}$. If $\mathbb{N}_{i_s}^{j_s}$ satisfies certain special properties, like \ac{SPC} and \ac{REP}, the fast decoding algorithms \cite{Alamdar2011Simp,Sarkis2014Fast,Hanif2017Fast,Condo2018Generalized,Zheng2021Threshold,Lu2024Fast,Sarkis2016Fast,Hashemi2017Fast,Ardakani2019Fast,Ren2022Sequence,Lu2025Fast} can directly obtain the \ac{ML} estimate or a list of candidates of the sub-codeword $\bm{s}^{N_s}$, without traversing the factor graph underneath the node. As such, the \ac{PM} update in (\ref{eq:PM_u}) is no longer applicable. Equivalently, \ac{PM} can be updated at the codeword side as \cite{hashemi2016fast}
\begin{equation}
    \PM^{(l)}_{j_s} = \PM^{(l)}_{i_s\s{-}1} + \sum_{k=1}^{N_s} \ln\left(1 + e^{-(1-2\hat{s}_l[k])\alpha^{(l)}_{\psi_s,\phi_s}[k]}\right),
\label{eq:PM_s}
\end{equation}
where $\hat{\bm{s}}_l^{N_s}$ is the estimated sub-codeword at the $l$-th path and $\bm{\alpha}^{(l)}_{\psi_s,\phi_s}$ is the corresponding internal \acp{LLR} passed to node $\mathbb{N}_{i_s}^{j_s}$. The estimate $\hat{\bm{s}}_l^{N_s}$ can either serve as the internal result $\bm{\beta}^{(l)}_{\psi_s,\phi_s}$ for subsequent decoding, or be used to obtain the estimate of $\bm{u}[i_s\s{:}j_s]$ by polar transform $\hat{\bm{u}}_l[i_s\s{:}j_s] = \bm{s}_l^{N_s} \bm{G}_{N_s}$.

\subsection{SO-SCL Decoding}
The extraction of bit-wise \acp{SO} in \ac{SO-SCL} decoding requires the assistance of a probability $P_\mathcal{T}(\bm{y}^N)$, which is defined as \cite{Yuan2025SoSCL}
\begin{equation}
    P_\mathcal{T}(\bm{y}^N) = \sum_{\bm{u}^N \in \mathcal{T}} P_{\bm{U}^N|\bm{Y}^N} \left(\bm{u}^N|\bm{y}^N\right),
\end{equation}
where $\mathcal{T}$ contains all valid input vectors $\bm{u}^N$ that satisfy the frozen constraints but are not included in the list output by \ac{SCL}. In \cite{Yuan2025SoSCL}, the authors proposed to approximate $P_\mathcal{T}(\bm{y}^N)$ by leveraging the \ac{SCL} decoding tree. 

\begin{figure}[!t]
    \centering
    \includegraphics[width=0.45\textwidth]{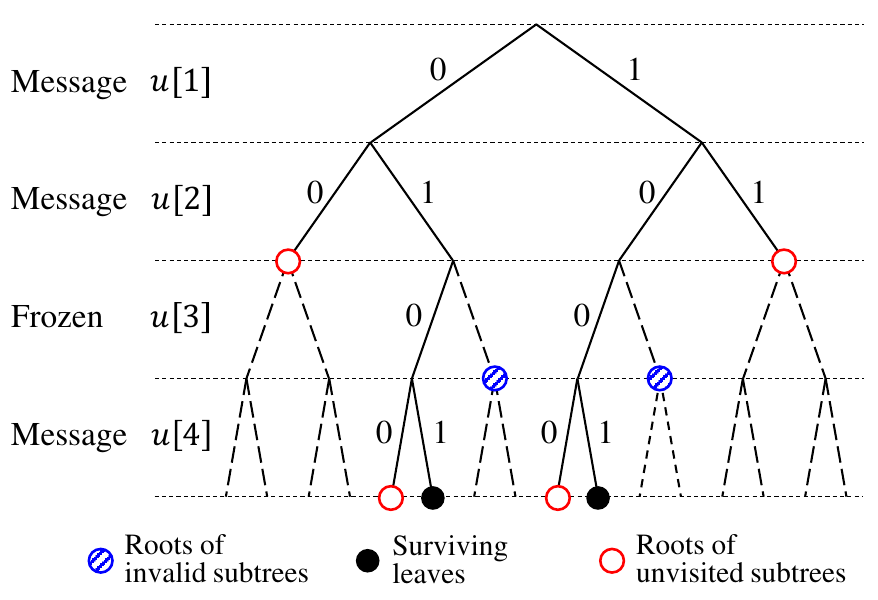}
    \caption{An example of the SCL decoding tree of a $(4,3)$ polar code with frozen bit $u[3]=0$ and list size $L=2$ \cite{Yuan2025SoSCL}. The whole tree consists of surviving leaves at $\mathcal{V}=\{(0,1,0,1), (1,0,0,1)\}$, invalid subtrees rooted at $\mathcal{B} =\{(0,1,1), (1,0,1)\}$, and unvisited subtrees rooted at $\mathcal{W}=\{(0,0), (1,1), (0,1,0,0), (1,0,0,0)\}$.}
    \label{fig:DecTree}
    \vspace{-10pt}
\end{figure}

An SCL decoding tree, illustrated in Fig. \ref{fig:DecTree}, consists of three parts \cite{Yuan2025SoSCL}: surviving leaves after \ac{SCL} decoding, unvisited valid subtrees, and invalid subtrees, where each node at the $i$-th level (root node is at the \mbox{$0$-th} level) corresponds to a possible decoding path $\bm{a}^i \in \{\bm{u}[1\s{:}i]:\bm{u}^N \in \{0, 1\}^N\}$ of depth $i$, and each leaf thus represents a possible input vector $\bm{u}^N$. The unvisited subtrees are pruned because of list size limitations, while the invalid subtrees are pruned due to conflict of frozen constraints. Let $\mathcal{V}$, $\mathcal{W}$, and $\mathcal{B}$ denote the sets of surviving leaves, roots of unvisited subtrees, and roots of invalid subtrees, respectively. 

Then, the probability $P_\mathcal{T}(\bm{y}^N)$ is approximated by \cite{Yuan2025SoSCL}
\begin{equation}
    P_\mathcal{T}^*(\bm{y}^N) = \underbrace{\sum\nolimits_{\bm{a}^i\in \mathcal{W}} 2^{-\left|\mathcal{F}^{(>i)}\right|} P_{\bm{U}^i|\bm{Y}^N}\left(\bm{a}^i|\bm{y}^N\right)}_{\text{Approx. sum of prob. for all unvisited valid leaves}} ,
\label{eq:Pt_appr}
\end{equation}
where $\mathcal{F}^{(>i)}$ contains frozen indices greater than $i$, defined by $\mathcal{F}^{(>i)} = \{j : j\in\mathcal{F}, i < j \leq N \}$. Given a (partial) decoding path $\bm{a}_l^{i}$ with path index $l$, the path probability $P_{\bm{U}^i|\bm{Y}^N}$ is associated with its \ac{PM} $\PM^{(l)}_{i}$ by \cite{Balats2015LLR}
\begin{equation}
    P_{\bm{U}^i|\bm{Y}^N} (\bm{a}_l^{i}|\bm{y}^N) = e^{-\PM^{(l)}_{i}}.
\label{eq:P_PM}
\end{equation}
Now, the bit-wise \ac{APP} \acp{LLR} $\ell_{\text{APP}}[i]$ are estimated by (\ref{eq:Lapp_appr}) at the bottom of the next page \cite{Yuan2025SoSCL}, where $\mathcal{U} = \mathcal{T} \cup \mathcal{V}$, $\mathcal{U}^{(i,b)}=\{\bm{c}^N: c[i]=b, \bm{c}^N=\bm{u}^N\bm{G}_N, \bm{u}^N\in\mathcal{U}\}$ and $\mathcal{V}^{(i,b)}=\{\bm{u}^N: c[i]=b, \bm{c}^N=\bm{u}^N\bm{G}_N, \bm{u}^N\in\mathcal{V}\}$.

\begin{figure*}[b]
\vspace{-10pt} \hrulefill
\begin{equation}
\begin{aligned}
    \ell_{\text{APP}}[i] &\triangleq \ln\frac{P_{C[i]|\bm{Y}^N}\left(0\,|\bm{y}^N\right) }{P_{C[i]|\bm{Y}^N}\left(1\,|\bm{y}^N\right) } 
    = \ln \frac{\sum_{\bm{u}^N\in\,\mathcal{U}^{(i,0)}} P_{\bm{U}^N|\bm{Y}^N}\left(\bm{u}^N|\bm{y}^N\right)}
    {\sum_{\bm{u}^N\in\,\mathcal{U}^{(i,1)}} P_{\bm{U}^N|\bm{Y}^N}\left(\bm{u}^N|\bm{y}^N\right)} \\
    &\approx \ln \frac{\sum_{\bm{u}^N\in\mathcal{V}^{(i,0)}} P_{\bm{U}^N|\bm{Y}^N}\left(\bm{u}^N|\bm{y}^N\right) + P_\mathcal{T}^*\left(\bm{y}^N\right) P_{C|Y}\left(0\,|y[i]\right)}
    {\sum_{\bm{u}^N\in\mathcal{V}^{(i,1)}} P_{\bm{U}^N|\bm{Y}^N}\left(\bm{u}^N|\bm{y}^N\right) + P_\mathcal{T}^*\left(\bm{y}^N\right) P_{C|Y}\left(1\,|y[i]\right)}, i\in[\![N]\!].
\end{aligned}
\label{eq:Lapp_appr}
\end{equation}
\vspace{-5pt}
\end{figure*}
\section{Proposed SO-FSCL Decoding}
To achieve lower decoding latency for \ac{SO-SCL}, we consider incorporating the fast decoding of some generalized special nodes. Observing that the \ac{FSCL} decoder can also output a list of candidate codewords like the conventional \ac{SCL} decoder, the key to \acp{SO} thus lies in calculating $P^*_{\mathcal{T}}(\bm{y}^N)$ for the considered nodes and using it to extract the bit-wise \ac{APP} \acp{LLR}. In this section, we begin with the node type constraint conducive to \ac{SO} generation. Subsequently, we investigate the estimation of $P^*_{\mathcal{T}}(\bm{y}^N)$ and its compatibility with dynamic frozen bits, followed by an improved \ac{APP} \ac{LLR} approximation method. Furthermore, based on the aforementioned probability-domain principles, we present the practical log-domain implementation and its \ac{HWF} variant, ultimately establishing the proposed \ac{SO-FSCL} decoding algorithm.

\subsection{Constraint on Node Types} \label{sec:3A}
Assuming that the node $\mathbb{N}_{i_s}^{j_s}$ corresponds to an $(N_s, K_s)$ sub-polar code, the frozen and information positions of the sub-polar code are indicated by $\mathcal{F}_s \subseteq \Idx{N_s}$ and $\mathcal{I}_s = \Idx{N_s} \cap \mathcal{F}_s^C$, respectively. In the proposed decoder, we constrain the node type to be in the form of \ac{FIC}, that is, $\mathcal{F}_s = \Idx{N_s \s{-} K_s}$, based on the following two considerations:
\begin{itemize}
    \item The approximation $P^*_{\mathcal{T}}(\bm{y}^N)$ is based on the assumption that $u[i]$ is uniformly distributed for $i \in \Idx{N}$, which implies that dynamic frozen bits are required \cite{Yuan2025SoSCL}. Since dynamic frozen bits are determined by previous bits, their computation becomes challenging if the frozen values depend on the undecoded information bits under the current node when applying fast decoding. Therefore, to universally support diverse frozen configurations, we adopt \ac{FIC} nodes to avoid this issue.
    \item The summation for calculating $P^*_{\mathcal{T}}(\bm{y}^N)$ requires the path probabilities of all nodes in $\mathcal{W}$. However, unlike conventional \ac{SCL} decoding, node-based fast decoding does not guarantee to access all these nodes \cite{Hashemi2017Fast,Ardakani2019Fast,Ren2022Sequence}, that is, cannot acquire some of their path probabilities. As we will elaborate in Section \ref{sec:3B}, the adopted \ac{FIC} nodes can significantly facilitate the estimation of path probabilities for those inaccessible nodes.
\end{itemize}

\begin{remark}
\label{mk:FIC}
    Under the node type constraint of $\mathcal{F}_s = \Idx{N_s - K_s}$, the number of information bits for the vast majority of \ac{FIC} nodes is limited by $\min\{K_s, N_s\s{-}K_s\} \leq 3$. For example, the 5G construction \cite{3GPP212} and beta-expansion construction \cite{he2017beta} inherently do not involve \ac{FIC} nodes with $\min\{K_s, N_s\s{-}K_s\} > 3$ according to their definitions. Meanwhile, as validated through our simulations, the Gaussian approximation construction \cite{Trif2012Effic} and Tal-Vardy construction \cite{tal2013construct} also rarely lead to such \ac{FIC} nodes. In the above case, \ac{FIC} nodes with $K_s = 0$, $1$, $2$, $3$, $N_s\s{-}3$, $N_s\s{-}2$, $N_s\s{-}1$ and $N_s$ correspond to the known \ac{Rate0}, \ac{REP}, Type-\Rmnum{1}, Type-\Rmnum{2}, Type-\Rmnum{4}, Type-\Rmnum{3}, \ac{SPC}, and \ac{Rate1} nodes, respectively \cite{Hashemi2017Fast,Ardakani2019Fast}.
\end{remark}

Note that since a polar code of length 2 must be an \ac{FIC} node, any polar code can be represented as a combination of several \ac{FIC} sub-polar codes. The set of \ac{FIC} nodes that constitute an $(N, K)$ polar code is denoted by $\mathcal{N}$ hereafter.

\subsection[PT(yN) Approximation for FIC Nodes]{$P^*_{\mathcal{T}}(\bm{y}^N)$ Approximation for FIC Nodes} \label{sec:3B}
For simplicity, we first assume all-zero frozen bits. After decoding a special node $\mathbb{N}_{i_s}^{j_s}$, the node-based \ac{FSCL} decoding algorithms, e.g. \cite{Hashemi2017Fast,Ardakani2019Fast,Ren2022Sequence,Lu2025Fast}, will directly jump from the ($i_s\s{-}1$)-th level to the $j_s$-th level in the decoding tree. If the set $\mathcal{V}_{i_s\s{-}1}$ contains the surviving nodes at the ($i_s\s{-}1$)-th level, there are $2^{K_s}|\mathcal{V}_{i_s\s{-}1}|$ valid descendant nodes at the $j_s$-th level. Among these, the set of nodes that remain surviving is denoted by $\mathcal{V}_{j_s}$, while the remaining nodes, i.e., the roots of pruned (unvisited) subtrees, constitute the set $\mathcal{W}_{j_s}$. When all nodes $\mathbb{N}_{i_s}^{j_s} \in \mathcal{N}$ are decoded, we have $\mathcal{V} = \mathcal{V}_{N}$ and $\mathcal{W} = \cup_{\mathbb{N}_{i_s}^{j_s}\in\mathcal{N}} \mathcal{W}_{j_s}$. 

Now, following the idea of the approximation (\ref{eq:Pt_appr}) in \ac{SO-SCL} decoding, the approximated probability $P^*_{\mathcal{T}}(\bm{y}^N)$ for \ac{FSCL} decoding is calculated as
\begin{equation}
    P^*_{\mathcal{T}}(\bm{y}^N) = \sum_{\mathbb{N}_{i_s}^{j_s}\in\mathcal{N}} P^*_\mathcal{W}(\mathbb{N}_{i_s}^{j_s}),
\label{eq:PT_appr}
\end{equation}
where the probability $P_\mathcal{W}^*(\mathbb{N}_{i_s}^{j_s})$ is defined by 
\begin{equation}
    P_\mathcal{W}^*(\mathbb{N}_{i_s}^{j_s}) = 2^{-\left|\mathcal{F}^{(>j_s)}\right|} \sum_{\bm{a}^{j_s} \in \mathcal{W}_{j_s}} \hspace{-5pt} P_{\bm{U}^{j_s}|\bm{Y}^N}\left(\bm{a}^{j_s}|\bm{y}^N\right).
\label{eq:PW}
\end{equation}
Hence, the computation of $P^*_{\mathcal{T}}(\bm{y}^N)$ essentially reduces to calculating $P_\mathcal{W}^*(\mathbb{N}_{i_s}^{j_s})$ after the hard-output \ac{FSCL} decoding of each node $\mathbb{N}_{i_s}^{j_s}$. Then, we will discuss the calculation of $P_\mathcal{W}^*(\mathbb{N}_{i_s}^{j_s})$ for \ac{FIC} nodes in two cases.

\begin{figure*}[!t]
    \centering
    \includegraphics[width=0.99\textwidth]{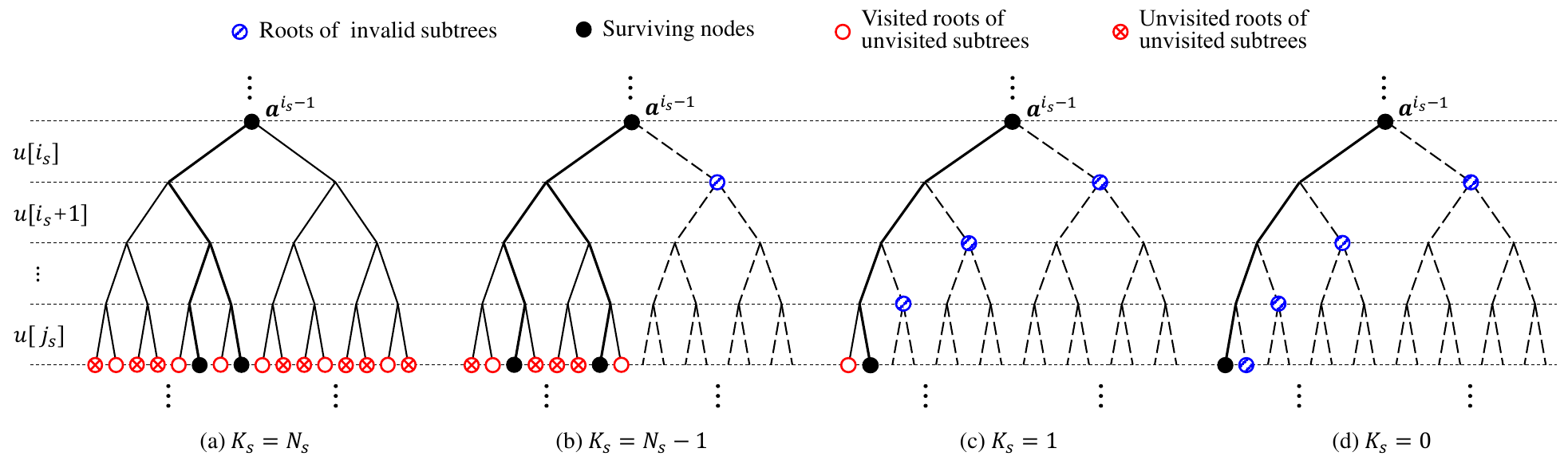}
    \caption{Examples of the partial \ac{FSCL} decoding tree for length-$N_s$ FIC nodes with different $K_s$ underneath a decoded path $\bm{a}^{i_s-1}$.}
    \label{fig:FSCLNode}
    \vspace{-10pt}
\end{figure*}

\subsubsection[Ks>Ns/2]{$K_s > N_s/2$} \label{sec:3B1}
When the code rate of \ac{FIC} nodes exceed 1/2, they could be regarded as a special case of SR1 nodes \cite{Lu2024Fast}, that is, the concatenation of a length-$N_q$ source node and several \ac{Rate1} nodes, where $N_q = \min\{x: x\in\{0, 2^0, 2^1, \cdots, \frac{N_s}{2}\}, x \geq N_s\s{-}K_s\}$. Then, a list of candidate sub-codewords need to be obtained through existing fast decoding algorithms. 

Specifically, if $N_q=0$, the \ac{FIC} nodes are equivalent to \ac{Rate1} nodes, and $L$ candidate sub-codewords can be efficiently searched via the bit-flipping based approach \cite{Hashemi2017Fast}. The decoding process initiates by performing hard decision on internal \acp{LLR} $\bm{\alpha}^{(l)}_{\psi_s,\phi_s}$ passed to $\mathbb{N}_{i_s}^{j_s}$ to obtain the \ac{ML} codewords for each path. Then, flip the least reliable bit in \ac{ML} codewords to double the paths, and $L$ ones with the lowest \acp{PM} survive. Repeat this step by flipping from the least reliable bit to the most reliable one for $\min\{L-1, N_s\}$ times. Moreover, if the source node is a \ac{Rate0} node, each subvector $\bm{q}_j = (s[j],s[j\s{+}N_q],\cdots,s[j\s{+}(\frac{N_s}{N_q}\s{-}1)N_q])$, $j\in\Idx{N_q}$, in sub-codeword $\bm{s}^{N_s}$ constitutes a \ac{SPC} code \cite{Condo2018Generalized}. The fast decoding of such \ac{FIC} nodes differs from \ac{Rate1} nodes in that the least reliable bit in each $\bm{q}_j$ should be determined aiming to perpetually satisfy even parity check and remains excluded from subsequent bit-flipping operations \cite{Condo2018Generalized,Ardakani2019Fast}. Through a straightforward extension of \cite[Theorem 1]{Ardakani2019Fast}, the number of required bit-flipping is $\min\{L-1, N_s\s{-}N_q\}$. Furthermore, when the source node is a \ac{REP} node, the parity checks on $\bm{q}_j$ can be either all odd or all even. Thus, we need to generate $2L$ \ac{ML} codewords for both hypotheses and retain the most probable $L$ codewords along with their corresponding parity-check constraints before bit-flipping, similar to the decoding of Type-\Rmnum{4} nodes in \cite{Ardakani2019Fast}. For more general cases, we may adopt the algorithm in \cite{Lu2025Fast} to perform fast list decoding. Please note that our work focuses on extracting \acp{SO} by leveraging the \ac{FSCL} decoding tree, rather than the hard-output fast decoding algorithms themselves. Hence, we refer readers to \cite{Hashemi2017Fast,Ardakani2019Fast,Lu2025Fast} for more detailed implementations of \ac{FSCL} decoding, including path splitting, path selection and \ac{PM} updating.

Since it may be computationally infeasible to traverse all $2^{K_s} |\mathcal{V}_{i_s\s{-}1}|$ valid sub-codewords, certain nodes in $\mathcal{W}_{j_s}$ will not be accessed by \ac{FSCL} decoding algorithms, resulting in their path probabilities being unknown. Figs. \ref{fig:FSCLNode}(a) and \ref{fig:FSCLNode}(b) illustrate two examples of this situation, where unvisited roots of unvisited subtrees are marked with crosses. Thus, directly calculating $P_\mathcal{W}^*(\mathbb{N}_{i_s}^{j_s})$ as in (\ref{eq:PW}) becomes challenging.

Nevertheless, by leveraging the \textit{fully-valid subtrees} in the decoding tree, we can derive the sum in (\ref{eq:PW}) at a higher level. Specifically, given a valid decoding path $\bm{a}^{i}$, let $\mathcal{D}_{j,\bm{a}^{i}}$ denote the set of all valid paths branching underneath it to the $j$-th level ($j>i$). If $\bm{u}[i\s{+}1:j]$ are all information bits, the subtree rooted at $\bm{a}^{i}$ with a depth of $j-i$ is termed fully-valid, satisfying $\mathcal{D}_{j,\bm{a}^{i}} = \{(\bm{a}^{i}, \bm{b}^{j-i}): \bm{b}^{j-i} \in \{0,1\}^{j-i}\}$. In this case, we have
\begin{equation}
\begin{aligned}
    \sum_{\bm{a}^{j} \in \mathcal{D}_{j,\bm{a}^{i}}} &P_{\bm{U}^{j}|\bm{Y}^N}\left(\bm{a}^{j}|\bm{y}^N\right) \\
    &\overset{(a)}{=} \sum_{\bm{b}^{j-i} \in \{0,1\}^{j-i}} \hspace{-5pt} P_{\bm{U}^{j}|\bm{Y}^N}\left((\bm{a}^{i}, \bm{b}^{j-i})|\bm{y}^N\right) \\
    &= P_{\bm{U}^{i}|\bm{Y}^N}\left(\bm{a}^{i}|\bm{y}^N\right).
\end{aligned}
\label{eq:FVtree}
\end{equation}
On the other hand, if $\bm{u}[i\s{+}1:j]$ contains any frozen bits, then $\mathcal{D}_{j,\bm{a}^{i}} \subset \{(\bm{a}^{i}, \bm{b}^{j-i}): \bm{b}^{j-i} \in \{0,1\}^{j-i}\}$, and consequently, the equality (a) in (\ref{eq:FVtree}) becomes a less-than sign. For an \ac{FIC} node $\mathbb{N}_{i_s}^{j_s}$, the deepest fully-valid subtrees branching to the \mbox{$j_s$-th} level are rooted at level $f_s$, where $f_s = j_s-K_s$. Since $\bm{u}[i_s:f_s]$ are all frozen, the set of valid nodes at level $f_s$ underneath the nodes in $\mathcal{V}_{i_s\s{-}1}$ can be represented as $\mathcal{D}_{f_s} = \{(\bm{a}^{i_s\s{-}1},\bm{0}^{N_s\s{-}K_s}): \bm{a}^{i_s\s{-}1}\in\mathcal{V}_{i_s\s{-}1}\}$. Furthermore, based on the fact that $\mathcal{V}_{j_s} \cup \mathcal{W}_{j_s} = \cup_{\bm{a}^{f_s} \in \mathcal{D}_{f_s}} \mathcal{D}_{j_s,\bm{a}^{f_s}} = \{(\bm{a}^{f_s},\bm{b}^{K_s}): \bm{a}^{f_s}\in\mathcal{D}_{f_s}, \bm{b}^{K_s}\in\{0,1\}^{K_s}\}$ and (\ref{eq:FVtree}), we obtain the following relationship
\begin{align}
    &\sum_{\bm{a}^{j_s} \in \mathcal{V}_{j_s}} \hspace{-5pt} P_{\bm{U}^{j_s}|\bm{Y}^N}\left(\bm{a}^{j_s}|\bm{y}^N\right) + \sum_{\bm{a}^{j_s} \in \mathcal{W}_{j_s}} \hspace{-5pt} P_{\bm{U}^{j_s}|\bm{Y}^N}\left(\bm{a}^{j_s}|\bm{y}^N\right) \notag \\
    \label{eq:observ}
    &= \sum_{\bm{a}^{f_s} \in \mathcal{D}_{f_s}} \hspace{2pt} \sum_{\bm{b}^{j_s} \in \mathcal{D}_{j_s,\bm{a}^{f_s}}} \hspace{-5pt} P_{\bm{U}^{j_s}|\bm{Y}^N}\left(\bm{b}^{j_s}|\bm{y}^N\right) \\
    &= \sum_{\bm{a}^{f_s} \in \mathcal{D}_{f_s}} \hspace{-5pt} P_{\bm{U}^{f_s}|\bm{Y}^N}\left(\bm{a}^{f_s}|\bm{y}^N\right). \notag
\end{align}
Eq. (\ref{eq:observ}) indicates that the calculation of $P_\mathcal{W}^*(\mathbb{N}_{i_s}^{j_s})$ in (\ref{eq:PW}) can be reduced from traversing all nodes in $\mathcal{W}_{j_s}$ to visiting only the nodes in $\mathcal{D}_{f_s}$, with a dimension of $|\mathcal{D}_{f_s}| = |\mathcal{V}_{i_s\s{-}1}| \leq L$, which is formulated as
\begin{equation}
\begin{aligned}
    P_\mathcal{W}^*(\mathbb{N}_{i_s}^{j_s}) = 2^{-\left|\mathcal{F}^{(>j_s)}\right|} \bigg( \sum_{\bm{a}^{f_s} \in \mathcal{D}_{f_s}} \hspace{-5pt} P_{\bm{U}^{f_s}|\bm{Y}^N}\left(\bm{a}^{f_s}|\bm{y}^N\right) \\
    - \sum_{\bm{a}^{j_s} \in \mathcal{V}_{j_s}} \hspace{-5pt} P_{\bm{U}^{j_s}|\bm{Y}^N}\left(\bm{a}^{j_s}|\bm{y}^N\right) \bigg).
\end{aligned}
\label{eq:PW_H}
\end{equation}

\begin{table}[t]
    \caption{Calculation of $P_\mathcal{W}^*(\mathbb{N}_{i_s}^{j_s})$ for FIC Nodes under Different Cases}
    \centering
    \begin{tabular}{@{~}cccc@{~}}
        \toprule
        \multirow{2}{*}{$K_s$} & \multicolumn{2}{c}{$\leq N_s/2$} & \multirow{2}{*}{$>N_s/2$} \\ \cmidrule(lr){2-3}
        & $\leq 3$ & $>3$ & \\ \midrule
        Calc. of $P_\mathcal{W}^*(\mathbb{N}_{i_s}^{j_s})$ & Eq. (\ref{eq:PW}) & Eq. (\ref{eq:PW_H}) & Eq. (\ref{eq:PW_H}) \\ \bottomrule
    \end{tabular}
    \label{tab:PW}
	\vspace*{-10pt}
\end{table}

If $K_s=N_s$, the probabilities $P_{\bm{U}^{f_s}|\bm{Y}^N}$ are directly given by the \acp{PM} from previous node decoding, whereas when $K_s<N_s$, we need to perform an additional \ac{SCL} decoder that only decodes the first $N_s-K_s$ frozen bits to obtain such probabilities. Since the frozen values are known, (\ref{eq:f_func}) and (\ref{eq:g_func}) can be computed in parallel down to the 0-th stage in the factor graph, disregarding the sequential nature of \ac{SCL} decoding. Moreover, the node-based fast decoding could also be applied here to further accelerate the computation. Observe that the high-rate \ac{FIC} nodes are mainly with $K_s = N_s\s{-}1$, $N_s\s{-}2$, or $N_s\s{-}3$ as noted in Remark~\ref{mk:FIC}, and $\bm{u}[i_s\s{:}f_s\s{+}1]$, $\bm{u}[i_s\s{:}f_s]$, and $\bm{u}[i_s\s{:}f_s{+}1]$ underneath these three types of nodes constitute \ac{REP}, \ac{Rate0}, and \ac{REP} nodes, respectively. We consider directly extracting the \ac{PM} of the last frozen bit when encountering a \ac{Rate0} or \ac{REP} node. Given a length-$N_s^\prime$ \ac{Rate0}/\ac{REP} node $\mathbb{N}_{i_s^\prime}^{j_s^\prime}$ and internal \acp{LLR} $\bm{\alpha}^{(l)}_{\psi_s^\prime,\phi_s^\prime}$ passed to it at the $l$-th path, the \ac{PM} $\PM^{(l)}_{f_s^\prime}$ ($f_s^\prime=j_s^\prime$) for \ac{Rate0} nodes is obtained by (\ref{eq:PM_s}), while the \ac{PM} $\PM^{(l)}_{f_s^\prime}$ ($f_s^\prime=j_s^\prime\s{-}1$) for \ac{REP} nodes is calculated by 
\begin{equation}
\begin{aligned}
    \PM^{(l)}_{f_s^\prime} =~ &\PM^{(l)}_{i_s^\prime\s{-}1} + \sum_{k=1}^{N_s^\prime} \ln\left(1 + e^{-\alpha^{(l)}_{\psi_s^\prime,
\phi_s^\prime}[k]}\right) \\
    &- \ln\left(1+e^{-\sum_{k=1}^{N_s^\prime}\alpha^{(l)}_{\psi_s^\prime,\phi_s^\prime}[k]}\right),
\end{aligned}
\end{equation}
which is derived from (\ref{eq:PM_s}), (\ref{eq:P_PM}), and the fact that
\begin{equation*}
\begin{aligned}
    P_{\bm{U}^{f_s^\prime}|\bm{Y}^N}&\big((\bm{a}^{
i_s^\prime\s{-}1}, \bm{0}^{N_s^\prime\s{-}1})|\bm{y}^N\big) = \\
    \sum_{b\in\{0,1\}} \hspace{-5pt} & P_{\bm{U}^{j_s^\prime}|\bm{Y}^N}\big((\bm{a}^{i_s^\prime\s{-}1}, \bm{0}^{N_s^\prime\s{-}1}, b)|\bm{y}^N\big), ~\bm{a}^{i_s^\prime\s{-}1} \in \mathcal{V}_{i_s^\prime\s{-}1}.
\end{aligned}
\end{equation*}
It is worth noting that the calculation of $P_{\bm{U}^{f_s}|\bm{Y}^N}$ is independent of the hard-output decoding process of the \ac{FIC} nodes, and thus can be carried out simultaneously.

\begin{remark}
    The basic idea behind our calculation of $P_\mathcal{W}^*(\mathbb{N}_{i_s}^{j_s})$ lies in utilizing fully-valid subtrees to merge the path probabilities of the nodes in $\mathcal{V}_{j_s} \cup \mathcal{W}_{j_s}$ up to the previous level. This concept is, in fact, applicable to any type of special nodes. Given a node $\mathbb{N}_{i_s}^{j_s}$, we redefine $f_s$ as the level of the last frozen bit within this node, and let the set $\mathcal{D}_{f_s}$ contain all valid nodes at the $f_s$-th level descending from the nodes in $\mathcal{V}_{i_s\s{-}1}$. In this case, we still preserve the relationship in (\ref{eq:observ}). However, unlike the \ac{FIC} structure that ensures no path splitting during the acquisition of path probabilities $P_{\bm{U}^{f_s}|\bm{Y}^N}$, the presence of $K_f = K_s - j_s + f_s$ information bits in $\bm{u}[i_s:f_s]$ causes the dimension of $\mathcal{D}_{f_s}$ to expand from $|\mathcal{V}_{i_s\s{-}1}|$ to $2^{K_f}|\mathcal{V}_{i_s\s{-}1}|$. This implies that the auxiliary \ac{FSCL} decoder incurs exponential complexity and storage costs to obtain the path probabilities for all nodes in $\mathcal{D}_{f_s}$. Although this overhead might be acceptable when $K_f$ is small, we must also consider compatibility with dynamic frozen bits. While it is possible to employ specific designs to make the frozen bits in node $\mathbb{N}_{i_s}^{j_s}$ independent of the information bits beneath it, taking into account the generality, implementability, and complexity of the decoder, our \ac{SO-FSCL} decoder is designed to identify only \ac{FIC} nodes.
\end{remark}

\subsubsection[Ks<=Ns/2]{$K_s \leq N_s/2$}
The \ac{FIC} nodes are a special case of the \ac{G-REP} nodes now \cite{Condo2018Generalized}, that is, the concatenation of several \ac{Rate0} nodes and a length-$N_p$ source node, where $N_p = \min\{x:x\in\{0, 2^0, 2^1, \cdots, \frac{N_s}{2}\}, x \geq K_s\}$. Following the algorithm proposed in \cite{Ren2022Sequence}, the fast list decoding of such \ac{FIC} nodes are divided into two phases: First, efficiently calculate the \acp{PM} for $L$ parent paths at the source node, and then perform list decoding for the source node. Since the source node is indeed an \ac{FIC} node investigated in Sec.~\ref{sec:3B1}, we can employ the same approach to decode it and compute $P_\mathcal{W}^*(\mathbb{N}_{i_s}^{j_s})$. Furthermore, for the most prevalent \ac{FIC} nodes with $K_s \leq 3$ highlighted in Remark~\ref{mk:FIC}, it is feasible to traverse all $2^{K_s} |\mathcal{V}_{i_s\s{-}1}|$ valid sub-codewords and retain the $L$ most probable ones, as is done for \ac{REP}, Type-\Rmnum{1}, and Type-\Rmnum{2} nodes \cite{Ardakani2019Fast}. In other words,  all nodes in $\mathcal{W}_{j_s}$ are visited, and two such examples are provided in Figs. \ref{fig:FSCLNode}(c) and \ref{fig:FSCLNode}(d). Hence, $P_\mathcal{W}^*(\mathbb{N}_{i_s}^{j_s})$ can be directly computed by (\ref{eq:PW}).

In short, to achieve the approximation of $P^*_{\mathcal{T}}(\bm{y}^N)$ under \ac{FIC} constraints, we calculate $P_\mathcal{W}^*(\mathbb{N}_{i_s}^{j_s})$ for the different cases of \ac{FIC} nodes, as summarized in Table 1, and then obtain $P^*_{\mathcal{T}}(\bm{y}^N)$ via summation in (\ref{eq:PT_appr}).

\subsection{Compatibility with Dynamic Frozen Bits}
When decoding a special node $\mathbb{N}_{i_s}^{j_s}$, the dynamic frozen bits $\hat{u}_{l}[i_s\s{-}1+i]$, $i \in \mathcal{F}_s$, for each path is required to be determined according to $\hat{\bm{u}}_{l}[1\s{:}i_s\s{-}2+i]$ beforehand. Then, we will discuss their impact on sub-codeword generation.

Let $\tilde{\bm{u}}_{l}^{N_s} = \hat{\bm{u}}_{l}[i_s\s{:}j_s]$ for clarity. We can decompose $\tilde{\bm{u}}_{l}^{N_s}$ into two vectors, $\tilde{\bm{u}}_{F,l}^{N_s}$ and $\tilde{\bm{u}}_{I,l}^{N_s}$, where $\tilde{\bm{u}}_{F,l}^{N_s}[\mathcal{F}_s] = \tilde{\bm{u}}_{l}^{N_s}[\mathcal{F}_s]$ and $\tilde{\bm{u}}_{I,l}^{N_s}[\mathcal{I}_s] = \tilde{\bm{u}}_{l}^{N_s}[\mathcal{I}_s]$, with the remaining positions set to zero. Thus, the estimated sub-codeword $\hat{\bm{s}}_{l}^{N_s}$ at the $l$-th path is represented as
\begin{equation}
    \hat{\bm{s}}_{l}^{N_s} = \tilde{\bm{u}}_{l}^{N_s} \bm{G}_{N_s} = \underbrace{\tilde{\bm{u}}_{F,l}^{N_s} \bm{G}_{N_s}}_{\hat{\bm{s}}_{F,l}^{N_s}} \oplus \underbrace{\tilde{\bm{u}}_{I,l}^{N_s} \bm{G}_{N_s}}_{\hat{\bm{s}}_{I,l}^{N_s}}.
\end{equation}
The term $\hat{\bm{s}}_{I,l}^{N_s} = \hat{\bm{s}}_{l}^{N_s} \oplus \hat{\bm{s}}_{F,l}^{N_s}$ is exactly the sub-codeword under the all-zero frozen assumption. Thus, based on the definition of \ac{LLR}, we first modify the internal \acp{LLR} $\bm{\alpha}^{(l)}_{\psi_s,\phi_s}$ passed to this node to 
\begin{equation}
    \tilde{\alpha}^{(l)}_{\psi_s,\phi_s}[j] = \left(1-2\hat{s}_{F,l}[j]\right) \alpha^{(l)}_{\psi_s,\phi_s}[j], j\in\Idx{N_s\s{-}K_s},
\label{eq:MdLLR}
\end{equation}
Then, we can apply the node decoding method introduced in Sec. \ref{sec:3B} to generate estimates $\hat{\bm{s}}_{I,l}^{N_s}$, and further obtain estimates $\hat{\bm{s}}_{l}^{N_s}$ under dynamic frozen conditions by
\begin{equation}
    \hat{s}_{l}[j] = \hat{s}_{I,l}[j] \oplus \hat{s}_{F,\eta[l]}[j], j\in\Idx{N_s\s{-}K_s},
\label{eq:dCW}
\end{equation}
where $\eta[l]$ indicates which path the current $l$-th path originated from before node decoding.

Generally, we should perform one matrix multiplication or polar encoding to compute $\hat{\bm{s}}_{F,l}^{N_s}$. This operation is entirely bit-wise and can be executed rapidly with very low resource overhead. Moreover, we can further avoid the encoding process by constraining the number of dynamic frozen bits. Specifically, we set only the first $F_d$ frozen bits (if have) in each \ac{FIC} node to be dynamic. The value of $F_d$ should be sufficiently small to enable finding the corresponding $\hat{\bm{s}}_{F,l}^{N_s}$ from $2^{F_d}$ possible sub-codewords via a look-up table, while not significantly degrading the performance of approximating $P^*_{\mathcal{T}}(\bm{y}^N)$. As such, we can immediately obtain $\hat{\bm{s}}_{F,l}^{N_s}$ according to $\tilde{\bm{u}}_{F,l}^{N_s}$.

\subsection{Bit-Wise SO Approximation}
Recall the estimation of the \ac{APP} \acp{LLR} in (\ref{eq:Lapp_appr}), we can extract the bit-wise \acp{SO} from the candidate list $\mathcal{V}$ and the probability $P^*_{\mathcal{T}}(\bm{y}^N)$ as
\begin{equation}
    \ell_{\text{APP}}[i] \approx \ln \frac{\sum_{\bm{u}^N\in\mathcal{V}^{(i,0)}} P_{\bm{U}^N|\bm{Y}^N}\left(\bm{u}^N|\bm{y}^N\right) + P_\mathcal{T}^*\left(\bm{y}^N\right) \s{\cdot} \Xi_i^0}
    {\sum_{\bm{u}^N\in\mathcal{V}^{(i,1)}} P_{\bm{U}^N|\bm{Y}^N}\left(\bm{u}^N|\bm{y}^N\right) + P_\mathcal{T}^*\left(\bm{y}^N\right) \s{\cdot} \Xi_i^1},
\label{eq:Lapp_apprF}
\end{equation}
for $i\in\Idx{N}$. According to the definition of \ac{APP} \acp{LLR}, the term $\Xi_i^b$ is essentially an estimate of
\begin{equation}
    \frac{\sum_{\bm{u}^N\in\mathcal{T}^{(i,b)}} P_{\bm{U}^N|\bm{Y}^N}\left(\bm{u}^N|\bm{y}^N\right)}{P_\mathcal{T}\left(\bm{y}^N\right)}, b\in\{0,1\},
\label{eq:Tb}
\end{equation}
where $\mathcal{T}^{(i,b)}$ is defined analogously to that of $\mathcal{V}^{(i,b)}$ in (\ref{eq:Lapp_appr}). In \ac{SO-SCL} decoding \cite{Yuan2025SoSCL}, $\Xi_i^b$ is directly set to
\begin{equation}
    \Xi_i^b = P_{C|Y}\left(b\,|y[i]\right).
\label{eq:Xi}
\end{equation}
However, $P_{C|Y}$ is a prior-decoding probability that may mislead the estimate of (\ref{eq:Tb}). For example, if all decoding paths in the list agree on $c[i]=0$, we intuitively have $\Xi_i^0 > \Xi_i^1$, whereas $P_{C|Y}$ does not necessarily satisfy $P_{C|Y}\left(0\,|y[i]\right) > P_{C|Y}\left(1\,|y[i]\right)$. Therefore, we use the list $\mathcal{V}$ to modify (\ref{eq:Xi}) as
\begin{equation}
    \Xi_i^b = \!\left\{\begin{aligned}
        &\max\!\left\{P_{C|Y}\left(0\,|y[i]\right), P_{C|Y}\left(1\,|y[i]\right)\right\}, &&\!\!\!\text{if}~\mathcal{V}^{(i,1\s{-}b)} = \emptyset; \\
        &\min\!\left\{P_{C|Y}\left(0\,|y[i]\right), P_{C|Y}\left(1\,|y[i]\right)\right\}, &&\!\!\!\text{if}~\mathcal{V}^{(i,b)} = \emptyset; \\
        &P_{C|Y}\left(b\,|y[i]\right), &&\!\!\!\text{otherwise}.
    \end{aligned}\right.
\label{eq:XiM}
\end{equation}
Since this modification is also feasible to the \ac{SO-SCL} decoding, it will be applied to both \ac{SO-SCL} and \ac{SO-FSCL} decoding by default unless otherwise specified in subsequent sections.

\subsection{Log-Domain and Its HWF Implementation}
For ease of exposition, the aforementioned principles of the \ac{SO-FSCL} decoding are introduced in the probability domain. However, since the probability-domain operations are prone to underflow \cite{Balats2015LLR}, leading to numerical instability, they are typically performed in the log-domain in practice. Therefore, in this subsection, we present the log-domain implementation of the proposed \ac{SO-FSCL} decoding and further propose a \ac{HWF} version to avoid nonlinear operations.

For the calculation of $P_\mathcal{T}^*\left(\bm{y}^N\right)$, we use a variable $\lambda_\mathcal{T}$ to keep track of the log-domain sum of $P^*_\mathcal{W}(\mathbb{N}_{i_s}^{j_s})$ in (\ref{eq:PT_appr}). After decoding each FIC node $\mathbb{N}_{i_s}^{j_s}$, we recursively update its value according to (\ref{eq:P_PM}), (\ref{eq:PW_H}), and (\ref{eq:PW}). Specifically, if $K_s > \min\{3, N_s/2\}$, we compute
\begin{equation}
\begin{aligned}
    \lambda_{\mathcal{T},j_s} = &\Big(\lambda_{\mathcal{T},i_s\s{-}1} - |\mathcal{F}_s|\ln2\Big) \boxplus \\
    &\left(\left(\underset{\bm{a}_l^{f_s} \in \mathcal{D}_{f_s}}{\bigboxplus} \hspace{-5pt} -\PM_{f_s}^{(l)}\right) \boxminus \left(\underset{\bm{a}_l^{j_s} \in \mathcal{V}_{j_s}}{\bigboxplus} \hspace{-5pt} -\PM_{j_s}^{(l)}\right)\right);
\end{aligned}
\label{eq:lambdaH}
\end{equation}
otherwise, we have
\begin{equation}
    \lambda_{\mathcal{T},j_s} = \Big(\lambda_{\mathcal{T},i_s\s{-}1} - |\mathcal{F}_s|\ln2\Big) \boxplus \left(\underset{\bm{a}_l^{j_s} \in \overline{\mathcal{W}}_{j_s}}{\bigboxplus} \hspace{-5pt} -\PM_{j_s}^{(l)}\right),
\label{eq:lambdaL}
\end{equation}
where $\lambda_{\mathcal{T},0}$ is initialized to $-\infty$ and the path index $l$ associates the decoding path $\bm{a}_l^{i}$ with its \ac{PM} $\PM_{i}^{(l)}$. We use $\overline{\mathcal{W}}_{j_s}$ to denote the set of discarded paths in the current path buffer to facilitate subsequent description. That is, if $K_s \leq \min\{3, N_s/2\}$, $\overline{\mathcal{W}}_{j_s}$ is still equivalent to $\mathcal{W}_{j_s}$; otherwise, it contains the discarded paths during the most recent bit-flipping. The operator $\boxplus$ and $\boxminus$ represent addition and subtraction of probabilities in log domain, respectively. They are defined as
\begin{equation}
\begin{aligned}
    x_1 \boxplus x_2 &= \ln \left(e^{x_1} + e^{x_2}\right) \\
    &= \max\{x_1, x_2\} + \ln\big(1+e^{-|x_1-x_2|}\big),
\end{aligned}
\end{equation}
and
\begin{equation}
\begin{aligned}
    x_1 \boxminus x_2 &= \ln \left(e^{x_1} - e^{x_2}\right) \\
    &= x_2 + \ln\left(e^{x_1-x_2} - 1\right), x_1 > x_2.
\end{aligned}
\end{equation}
Hence, $P^*_{\mathcal{T}}(\bm{y}^N)$ in log domain is expressed as
\begin{equation}
    \ln P^*_{\mathcal{T}}(\bm{y}^N) = \lambda_{\mathcal{T},N}.
\label{eq:lambdaT}
\end{equation}
By (\ref{eq:Lapp_apprF}) and (\ref{eq:XiM}), the \ac{APP} \acp{LLR} are then estimated by (\ref{eq:Lapp_apprL}) at the bottom of this page.
\begin{figure*}[b]
\vspace{-10pt} \hrulefill
\begin{align}
    \label{eq:Lapp_apprL}
    \ell_{\text{APP}}[i] &\approx \left\{\begin{aligned}
    &\left(\bigboxplus_{\bm{u}_l^N \in \mathcal{V}} -\PM_N^{(l)}\right) \boxplus \left(\lambda_{\mathcal{T},N} + \max\{\lambda_{0|Y_i},\lambda_{1|Y_i}\}\right) - \left(\lambda_{\mathcal{T},N} + \min\{\lambda_{0|Y_i},\lambda_{1|Y_i}\}\right), && \text{if}~\mathcal{V}^{(i,1)} = \emptyset; \\
    &\left(\lambda_{\mathcal{T},N} + \min\{\lambda_{0|Y_i},\lambda_{1|Y_i}\}\right) - \left(\bigboxplus_{\bm{u}_l^N \in \mathcal{V}} -\PM_N^{(l)}\right) \boxplus \left(\lambda_{\mathcal{T},N} + \max\{\lambda_{0|Y_i},\lambda_{1|Y_i}\}\right), && \text{if}~\mathcal{V}^{(i,0)} = \emptyset; \\
    &\left(\bigboxplus_{\bm{u}_l^N \in \mathcal{V}^{(i,0)}} -\PM_N^{(l)}\right) \boxplus \left(\lambda_{\mathcal{T},N} + \lambda_{0|Y_i}\right) - \left(\bigboxplus_{\bm{u}_l^N \in \mathcal{V}^{(i,1)}} -\PM_N^{(l)}\right) \boxplus (\lambda_{\mathcal{T},N} + \lambda_{1|Y_i}), &&\text{otherwise}, \\
    \end{aligned}\right. \\
    &\text{where}\quad \lambda_{0|Y_i} = \ln P_{C|Y}\left(0\,|y[i]\right) = -\ln(1+e^{-\ell_{\rm ch}[i]}), \lambda_{1|Y_i} = \ln P_{C|Y}\left(1\,|y[i]\right) = \lambda_{0|Y_i} - \ell_{\rm ch}[i].\notag
\end{align}
\vspace{-5pt}
\end{figure*}

Notably, the \ac{SO-FSCL} decoding involves two nonlinear functions --- $\ln(1+e^x)$ and $\ln(e^x-1)$, which poses significant challenges for hardware implementation. Therefore, we attempt to avoid such nonlinear operations in the \ac{HWF} version of \ac{SO-FSCL} decoding. 

The function $\ln(1+e^x)$, also known as the \textit{softplus} function, can typically be approximated by linear/polynomial functions or lookup tables \cite[Sec. \Rmnum{5}-G]{Rowshan2024Channel}. For example, we may adopt the following approximation:
\begin{equation}
    \ln(1+e^x) \approx \left\{\begin{aligned}
        &\max\{0, \frac{1}{4}x + \ln2\}, && \text{if}~ x \leq 0; \\
        &\max\{x, \frac{3}{4}x + \ln2\}, && \text{otherwise},
    \end{aligned}\right.
\label{eq:jaclog}
\end{equation}
where the approximation for $x \leq 0$ originates from \cite{Jung2000Linearly}, while the approximation for the remaining interval is derived from the equation $\ln(1+e^x) - (1+e^{-x}) = x$. 

As for the function $\ln(e^x-1)$, we reformulate (\ref{eq:lambdaH}) to eliminate such nonlinear computation as follows:
\begin{equation}
\begin{aligned}
    \lambda_{\mathcal{T},j_s} \approx & \Big(\lambda_{\mathcal{T},i_s\s{-}1} - |\mathcal{F}_s|\ln2\Big) \boxplus \left(\underset{\bm{a}_l^{f_s} \in \mathcal{D}_{f_s\backslash j_s}}{\bigboxplus} \hspace{-5pt} -\PM_{f_s}^{(l)}\right) \\
    & \boxplus \left(\underset{\bm{a}_l^{j_s} \in \overline{\mathcal{W}}_{j_s}}{\bigboxplus} \hspace{-5pt} \ln(L-1)-\PM_{j_s}^{(l)}\right),
\end{aligned}
\label{eq:lambdaHWF}
\end{equation}
where the set $\mathcal{D}_{f_s\backslash j_s}$ is obtained by excluding from $\mathcal{D}_{f_s}$ the decoding paths that are inherited by any path in $\mathcal{V}_{j_s}$. Eq. (\ref{eq:lambdaHWF}) avoids the $\boxminus$ operation through a naive estimation of the sum of path probabilities for nodes in $\overline{\mathcal{W}}_{j_s}$. Thus, the \ac{HWF} version will only contain the softplus function, and this function can be approximated with much lower computational cost.

\subsection{Summary of SO-FSCL Decoding}
\begin{figure}[!t]
    \centering
    \includegraphics[width=0.48\textwidth]{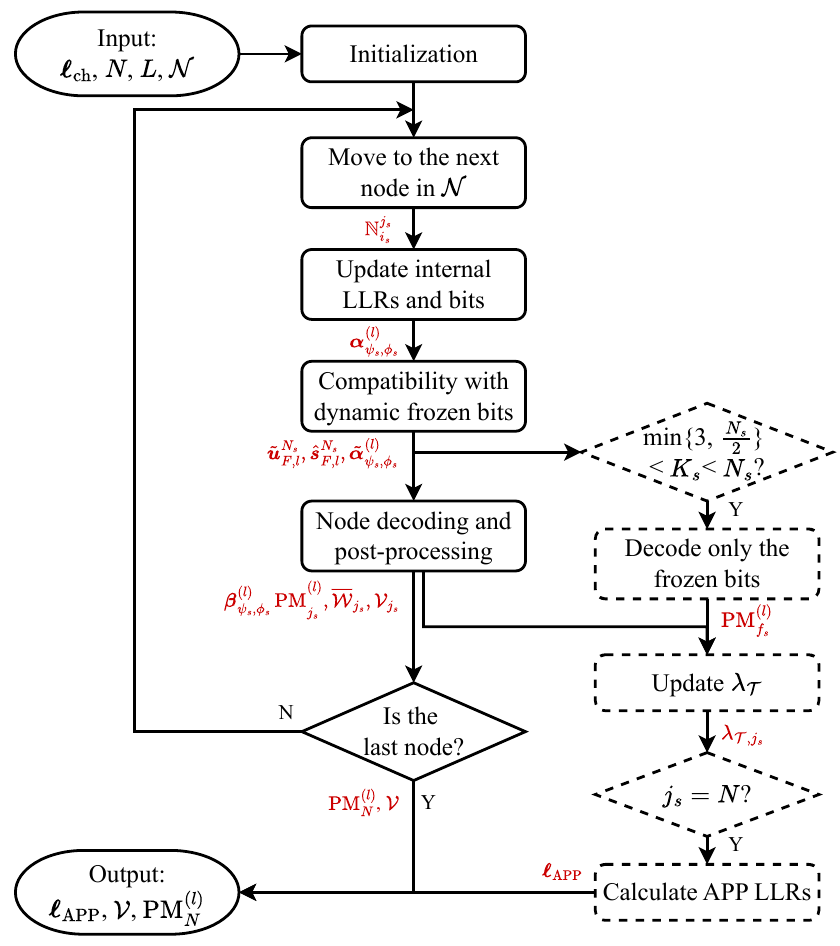}
    \caption{Flowchart of SO-FSCL Decoding. The dashed blocks represent modules related to SO extraction.}
    \label{fig:SO-FSCL}
    \vspace{-10pt}
\end{figure}

In Algorithm \ref{alg:SO-FSCL}, we summarize the procedure of proposed log-domain \ac{SO-FSCL} decoding. To simplify the notation, we use $\bm{\alpha}^{(l)}$ to represent all $\bm{\alpha}_{\psi,\phi}^{(l)}$ where $\psi \in \Idx{0\s{:}n}$ and $\phi \in \Idx{2^{n-\psi}}$, and the same applies to $\bm{\beta}^{(l)}$. Meanwhile, $\{\bm{\alpha}^{(l)}\}_{\mathcal{V}_0}$ denotes the set of $\bm{\alpha}^{(l)}$ corresponding to the $l$-th path in $\mathcal{V}_0$, and similar definitions are adopted for other variable sets related to path indices. The functions ${\rm UpdLLR}()$ and ${\rm NodeDec}()$ obtain the internal \acp{LLR} passed to the current node and perform node decoding as in conventional hard-output \ac{FSCL} decoder, while the function ${\rm FrzDec}()$ executes the auxiliary \ac{FSCL} decoder that decodes only the frozen bits as detailed in Sec. \ref{sec:3B}. For a more clearer demonstration of the decoding process, the flowchart of the \ac{SO-FSCL} decoding is further illustrated in Fig. \ref{fig:SO-FSCL}. It is observed that \ac{SO-FSCL} decoding can be regarded as an add-on extension to \ac{FSCL} decoding, that is, the \ac{SO} extraction depends on but operates independently of \ac{FSCL} decoding. Therefore, we hold the flexibility to configure \ac{SO-FSCL} decoder to output only hard decisions like \ac{FSCL} decoder or generate additional \acp{SO}.

\begin{algorithm}[!t]
    \caption{Log-Domain SO-FSCL Decoding}
    \label{alg:SO-FSCL}
    \begin{algorithmic}[1]
        \small
        \Require Channel LLRs $\bm{\ell}_{\rm ch}$, code length $N$, list size $L$, and FIC node set $\mathcal{N}$ (ascending order of $i_s$).

        \State $\mathcal{V}_0 \gets \{\bm{a}^0\}$; $\{\bm{\alpha}^{(l)}\}_{\mathcal{V}_0} \gets 0$; $\{\bm{\beta}^{(l)}\}_{\mathcal{V}_0} \gets 0$; $\{\bm{\alpha}_{n,1}^{(l)}\}_{\mathcal{V}_0} \gets \bm{\ell}_{\rm ch}$;

        \Statex\hspace{0.5em} $\{\PM_0^{(l)}\}_{\mathcal{V}_0} \gets 0$; $\lambda_{\mathcal{T}, 0} \gets -\infty$; \vspace{0.1cm}

        \For {$k=1$ \textbf{to} $|\mathcal{N}|$}
            \State $\mathbb{N}_{i_s}^{j_s} \gets k$-th node in $\mathcal{N}$; 
            
            \vspace{0.2cm} \Statex\hspace{1em} \texttt{// Update internal LLRs and bits} \vspace{0.05cm}
            
            \State $\{\bm{\alpha}^{(l)}_{\psi_s,\phi_s}\!\}_{\mathcal{V}_{i_s\s{-}1}} \s{\gets} {\rm UpdLLR}\big(\mathbb{N}_{i_s}^{j_s}, \{\bm{\alpha}^{(l)}\}_{\mathcal{V}_{i_s\s{-}1}}, \{\bm{\beta}^{(l)}\}_{\mathcal{V}_{i_s\s{-}1}}\big)$;
        
            \vspace{0.2cm} \Statex\hspace{1em} \texttt{// Compatibility with dynamic frozen bits} \vspace{0.05cm}

            \State Generate $\{\tilde{\bm{u}}_{F,l}^{N_s}\}_{\mathcal{V}_{i_s\s{-}1}}$ and corresponding $\{\hat{\bm{s}}_{F,l}^{N_s}\}_{\mathcal{V}_{i_s\s{-}1}}$;
            \State $\{\tilde{\bm{\alpha}}^{(l)}_{\psi_s,\phi_s}\!\}_{\mathcal{V}_{i_s\s{-}1}} \gets$ Modify $\{\bm{\alpha}^{(l)}_{\psi_s,\phi_s}\!\}_{\mathcal{V}_{i_s\s{-}1}}$ by (\ref{eq:MdLLR});

            \vspace{0.2cm} \Statex\hspace{1em} \texttt{// Hard-output node decoding} \vspace{0.05cm}

            \State $\mathcal{V}_{j_s}$, $\overline{\mathcal{W}}_{j_s}$, $\{\PM_{j_s}^{(l)}\}_{\mathcal{V}_{j_s} \cup \overline{\mathcal{W}}_{j_s}}$, $\{\hat{\bm{s}}_{I,l}^{N_s}\}_{\mathcal{V}_{j_s}}$, $\bm{\eta}$ $\gets$
            \Statex\hspace{2em} ${\rm NodeDec}\big(\mathbb{N}_{i_s}^{j_s}, \mathcal{V}_{i_s\s{-}1}, \{\tilde{\bm{\alpha}}^{(l)}_{\psi_s,\phi_s}\!\}_{\mathcal{V}_{i_s\s{-}1}}, \{\PM_{i_s\s{-}1}^{(l)}\!\}_{\mathcal{V}_{i_s\s{-}1}}, L\big)$;

            \vspace{0.2cm} \Statex\hspace{1em} \texttt{// Post-decoding processing} \vspace{0.05cm}

            \State $\{\bm{\beta}^{(l)}_{\psi_s,\phi_s}\}_{\mathcal{V}_{j_s}} \gets$ Calculate $\{\hat{\bm{s}}_{l}^{N_s}\}_{\mathcal{V}_{j_s}}$ by (\ref{eq:dCW});

            \State $\bm{a}_l[\mathcal{F}_s \s{+} i_s \s{-} 1] \gets \tilde{\bm{u}}_{F,\eta[l]}[\mathcal{F}_s]$, for all $\bm{a}_l^{j_s}$ \textbf{in} $\mathcal{V}_{j_s}$;

            \vspace{0.15cm} \Statex\hspace{1em} \texttt{// Decode only the frozen bits in $\mathbb{N}_{i_s}^{j_s}$} \vspace{0.05cm}

            \If{$\min\{3, N_s/2\}<K_s<N_s$}\vspace{0.02cm}
                \State $\{\PM_{f_s}^{(l)}\}_{\mathcal{V}_{f_s}} \gets$ 
                
                \Statex\hspace{3.5em} ${\rm FrzDec}\big(\mathbb{N}_{i_s}^{j_s}, \{\tilde{\bm{\alpha}}^{(l)}_{\psi_s,\phi_s}\!\}_{\mathcal{V}_{i_s\s{-}1}}, \{\PM_{i_s\s{-}1}^{(l)}\!\}_{\mathcal{V}_{i_s\s{-}1}}\big)$;
            \EndIf

            \vspace{0.15cm} \Statex\hspace{1em} \texttt{// Update $\lambda_\mathcal{T}$} \vspace{0.05cm}

            \If{$K_s>\min\{3, N_s/2\}$}
                \State Update $\lambda_{\mathcal{T}, j_s}$ by (\ref{eq:lambdaHWF}) for HWF, and (\ref{eq:lambdaH}) otherwise;
            \Else
                \State Update $\lambda_{\mathcal{T}, j_s}$ by (\ref{eq:lambdaL});
            \EndIf
        \EndFor

        \State $\mathcal{V} \gets \mathcal{V}_N$;
        \State Calculate $\bm{\ell}_{\rm APP}$ according to (\ref{eq:Lapp_apprL});
        \Ensure APP LLRs $\bm{\ell}_{\rm APP}$, candidate list $\mathcal{V}$, and \acp{PM} $\{\PM_N^{(l)}\}_\mathcal{V}$.
    \end{algorithmic}
\end{algorithm}

\section{Decoding Latency and Complexity Analysis}
\subsection{Decoding Latency with Unlimited Resources}
The decoding latency can be evaluated by the required number of time steps to decode a polar code. We adopt the following assumptions used in \cite{Hanif2017Fast,Condo2018Generalized,Zheng2021Threshold,Lu2024Fast,Hashemi2017Fast,Ardakani2019Fast,Lu2025Fast,Shen2022Fast,hashemi2016fast} for analyzing decoding latency: 1) we assume that there is no resource limitation for operations that can be executed in parallel, 2) basic operation of real numbers and check-node operation require one time step, 3) hard decisions, bit operations, and sign operations can be carried out instantaneously, 4) path splitting, sorting of $2^k L$ PMs and selection of the most probable $L$ paths consume $k$ time steps ($k=1,2,3,\cdots$), and 5) it takes one time step to obtain the \ac{ML} codeword of a \ac{SPC} node. Furthermore, the softplus function is involved in $f$-function computation, \ac{PM} update, and boxplus $\boxplus$ operation throughout the decoding process. By applying the \ac{HWF} version, we assume that each of the above three operations consumes one time step.

The required time steps for \ac{SO-FSCL} decoding primarily come from decoding special nodes and calculating \ac{SO} related values. At the beginning of decoding an \ac{FIC} node $\mathbb{N}_{i_s}^{j_s}$, we need to handle the compatibility with dynamic frozen bits, including determining their values and modifying the internal \acp{LLR} accordingly. Since this step only involves bit and sign operations, it is immediately carried out. Then, an \ac{FSCL} decoder requires $T_h$ time steps to perform the hard-output decoding of the \ac{FIC} node. Meanwhile, if $\min\{3, N_s/2\} < K_s < N_s$, we consume $T_p$ time steps to parallelly decode only the frozen bits in node $\mathbb{N}_{i_s}^{j_s}$. According to the method described in Sec. \ref{sec:3B1}, we have $T_p = \log_2(N_s/N_m) + 1$, where $N_m$ is the minimum length of the identified \ac{Rate0} and \ac{REP} nodes. Subsequently, updating the variable  $\lambda_{\mathcal{T}, j_s}$ takes $T_o = \left\lceil\log_2(2^{K_s}L - L + 1)\right\rceil $ time steps for $0 < K_s \leq \min\{3, N_s/2\}$ and at most $T_o = \left\lceil\log_2 2L\right\rceil$ time steps for $K_s > \min\{3, N_s/2\}$, where $\left\lceil \cdot\right\rceil$ indicates the ceiling function. Clearly, $T_o = 0$ when $K_s = 0$. Since updating $\lambda_{\mathcal{T}, j_s}$ can proceed in parallel with the decoding of the next node, the required time steps to decode an \ac{FIC} node are expressed as $T_s = \max\{T_h, T_p, T_o^\prime\s{-}T_l\}$, where $T_o^\prime$ denotes $T_o$ of the previous node and $T_l$ refers to the time steps for computing internal \acp{LLR} passed to current node. Finally, the proposed \ac{SO-FSCL} decoder estimates the \ac{APP} \acp{LLR} within $1+\left\lceil\log_2 L\right\rceil$ time steps.

In Table \ref{tab:ts_node}, we provide the required number of time steps to decode the most prevalent \ac{FIC} nodes. For comparison, we present the node decoding performance of \ac{FSCL}, \ac{SO-FSCL}, \ac{SO-SCL}, and \ac{FastSOL} decoders. The \ac{FSCL} decoder, which also serves as the inner hard-output decoder in \ac{SO-FSCL}, identifies the same special nodes as in \ac{SO-FSCL} and employs the algorithms proposed in \cite{Ardakani2019Fast} for node decoding\footnote{In the latency analysis, complexity analysis, and simulations of this paper, the FSCL decoder employs such configuration throughout.}. Since \ac{SO-SCL} relies on the conventional \ac{SCL} decoder \cite{Yuan2025SoSCL}, it takes $2N_s\s{+}K_s\s{-}2$ time steps to generate the hard-output for an $(N_s, K_s)$ sub-polar code \cite{Hashemi2017Fast}. Similar to \ac{SO-FSCL}, the update of (\ref{eq:Pt_appr}) can be done in parallel with subsequent \ac{SCL} decoding. Moreover, \ac{SO-SCL} also consumes $1+\left\lceil\log_2 L\right\rceil$ time steps to calculate the \ac{APP} \acp{LLR}. In \ac{FastSOL} decoder \cite{Shen2022Fast}, only three types of nodes are identified: \ac{Rate0}, \ac{Rate1} and \ac{REP}, with their \ac{SO} node decoding requiring 1, 3 and 2 time steps respectively. Furthermore, \ac{FastSOL} takes an additional time step to perform one message passing for child nodes to their parent nodes, because it is \ac{SCAN}-based and equations (\ref{eq:b1_func}) and (\ref{eq:b2_func}) are replaced by real-valued operations that include one $f$-function \cite{Shen2022Fast}. 

\begin{table*}
    \caption{Required Number of Time Steps to Decode the Most Prevalent FIC nodes of Length $N_s$ and List Size $L$}
	\centering
    \renewcommand{\arraystretch}{1.2}
    \setlength{\tabcolsep}{4.5pt}
	\begin{tabular}{@{~}llcccccccc@{~}}
		\toprule
		& & $K_s=0$ & $K_s=1$ & $K_s=2$ & $K_s=3$ & $K_s=N_s\s{-}3$ & $K_s=N_s\s{-}2$ & $K_s=N_s\s{-}1$ & $K_s=N_s$ \\
        [1pt]\toprule
        FSCL & $T_h$ & 1 & 2 & 3 & 4 & $\min\{L, N_s\s{-}3\}\s$ & $\min\{L, N_s\s{-}1\}$ & $\min\{L, N_s\}$ & $\min\{L, N_s\s{+}1\}$ \\
        \midrule
        & $T_p$ & 0 & 0 & 0 & 0 & $\log_2 (N_s/2)$ & $\log_2 N_s$ & $\log_2 N_s$ & 0 \\ \addlinespace[1pt]
        SO-FSCL & $T_o$ & 0 & $\left\lceil\log_2 (L\s{+}1)\right\rceil$ & $\left\lceil\log_2 (3L\s{+}1)\right\rceil$ & $\left\lceil\log_2 (7L\s{+}1)\right\rceil$ & $\left\lceil\log_2 2L\right\rceil$ & $\left\lceil\log_2 2L\right\rceil$ & $\left\lceil\log_2 2L\right\rceil$ & $\left\lceil\log_2 2L\right\rceil$ \\
        \cmidrule(l){2-10}
        & $T_s$ & \multicolumn{8}{c}{$\max\{T_h, T_p, T_o^\prime-T_l\}$} \\
        \midrule
        SO-SCL & & $2N_s\s{-}2$ & $2N_s\s{-}1$ & $2N_s\s$ & $2N_s\s{+}1$ & $3N_s\s{-}5$ & $3N_s\s{-}4$ & $3N_s\s{-}3$ & $3N_s\s{-}2$ \\
        \midrule
        FastSOL & & 1 & 2 & $4\log_2N_s\s{-}1$ & $4\log_2N_s$ & $6\log_2N_s\s{-}10$ & $6\log_2N_s\s{-}5$ & $6\log_2N_s\s{-}4$ & 3 \\
		\bottomrule
	\end{tabular}
	\label{tab:ts_node}
	\vspace*{-10pt}
\end{table*}

\begin{table}
    \caption{Required Number of Time Steps to Decode Polar Codes with Different Polar Decoders}
    \centering
    \renewcommand{\arraystretch}{1.1}
    \begin{tabular}{@{~}lccccc@{~}}
        \toprule
        $(N,K)$ & $L$ & SO-SCL & SO-FSCL & FastSOL & FSCL \\
        [1pt]\toprule
        & 2 & 598 & 83 & 193 & 74 \\
        (256,85) & 4 & 600 & 97 & 193 & 90 \\
        & 8 & 636 & 116 & 193 & 106 \\
        \midrule
        & 2 & 1281 & 173 & 398 & 155 \\
        (512,256) & 4 & 1283 & 201 & 398 & 193 \\
        & 8 & 1399 & 248 & 398 & 240 \\
        \midrule
        & 2 & 2561 & 292 & 714 & 260 \\
        (1024,512) & 4 & 2563 & 338 & 714 & 326 \\
        & 8 & 2794 & 415 & 714 & 405 \\
        \bottomrule
	\end{tabular}
    \label{tab:ts_polar}
	\vspace*{-10pt}
\end{table}

Then, we count the required time steps for the aforementioned four decoders to decode polar codes with different code lengths and code rates, as shown in Table \ref{tab:ts_polar}. Polar codes are constructed in accordance with the 5G standard \cite{3GPP212}, with their \ac{FIC} node compositions listed in Table \ref{tab:node_compo}. It is observed in Table \ref{tab:ts_polar} that the proposed \ac{SO-FSCL} decoder can save at least 81.8\% of the time steps with respect to the \ac{SO-SCL} decoder. By identifying more generalized nodes, \ac{SO-FSCL} decoding remains faster than \ac{FastSOL} decoding, with time step reduction being at least 37.7\%. However, compared to \ac{FSCL} decoder, \ac{SO-FSCL} decoder needs to generate \acp{SO} at the cost of a little decoding latency. Note that the reduction in decoding latency attributed to node-based decoding is related only to the structure of polar codes and is independent of the channel conditions. Furthermore, Table \ref{tab:node_compo} confirms that the prevalent \ac{FIC} nodes satisfy $\min\{K_s, N_s\s{-}K_s\} \leq 3$ as annotated in Remark \ref{mk:FIC}.

\begin{table}
    \caption{Number of FIC Nodes Constituting Different Polar Codes}
    \centering
    \renewcommand{\arraystretch}{1}
    \setlength{\tabcolsep}{4pt} 
    \begin{tabular}{@{~}lccccccccc@{~}}
        \toprule
        \multirow{2}{*}[-2pt]{$(N,K)$} & \multirow{2}{*}[-2pt]{$N_s$}~~ & \multicolumn{8}{c}{$K_s$} \\
        \cmidrule(l){3-10}
        & & ~0~ & ~1~ & ~2~ & ~3~ & $N_s\s{-}3$ & $N_s\s{-}2$ & $N_s\s{-}1$ & $N_s$ \\
        [1pt]\toprule
        \multirow{5}{*}{(256,85)} & 4~~ & 0 & 4 & 0 & 0 & 0 & 0 & 4 & 0 \\
        & 8~~ & 0 & 2 & 0 & 0 & 0 & 0 & 1 & 1 \\
        & 16~~ & 0 & 0 & 0 & 1 & 1 & 0 & 0 & 0 \\
        & 32~~ & 0 & 1 & 0 & 1 & 0 & 0 & 1 & 0 \\
        & 64~~ & 0 & 1 & 0 & 0 & 0 & 0 & 0 & 0 \\
        \midrule
        \multirow{5}{*}{(512,256)} & 4~~ & 0 & 7 & 0 & 0 & 0 & 0 & 7 & 0 \\
        & 8~~ & 2 & 3 & 1 & 1 & 0 & 1 & 3 & 2 \\
        & 16~~ & 1 & 1 & 1 & 0 & 0 & 1 & 1 & 1 \\
        & 32~~ & 0 & 2 & 0 & 0 & 0 & 0 & 2 & 0 \\
        & 64~~ & 0 & 1 & 0 & 0 & 0 & 0 & 0 & 1 \\
        \midrule
        \multirow{6}{*}{(1024,512)} & 4~~ & 0 & 11 & 0 & 0 & 0 & 0 & 11 & 0 \\
        & 8~~ & 2 & 6 & 1 & 1 & 3 & 0 & 7 & 1 \\
        & 16~~ & 1 & 4 & 1 & 0 & 0 & 1 & 4 & 1 \\
        & 32~~ & 1 & 1 & 0 & 1 & 1 & 0 & 0 & 2 \\
        & 64~~ & 0 & 1 & 0 & 0 & 0 & 0 & 1 & 0 \\
        & 128~~ & 0 & 1 & 0 & 0 & 0 & 0 & 1 & 0 \\
        \bottomrule
	\end{tabular}
    \label{tab:node_compo}
	\vspace*{-10pt}
\end{table}
\subsection{Decoding Latency with Resource Constraints}
In case practical hardware implementations cannot offer unlimited computational resources, the latency performance of the proposed \ac{SO-FSCL} decoder under resource constraints is also evaluated. We assume that only $N_{\rm par}$ operations can be executed parallelly at the same time, which indicates that if the number of parallel operations exceeds $N_{\rm par}$, additional time steps will be required. Moreover, we stipulate that the operations for hard outputs and \acp{SO} are equally allocated resources, that is, each will occupy half of the parallel resources at full load. In Table \ref{tab:ts_limit}, we further present the variation of decoding latency with $N_{\rm par}$. At low parallelism levels, such as $N_{\rm par} = 2L$, the latency savings of \ac{SO-FSCL} compared to \ac{SO-SCL} decrease from at least 81.8\% to approximately 54\%, while the latency improvement over \ac{FSCL} also shows a slight increase. However, as $N_{\rm par}$ increases from $2L$ to $32L$, the performance of \ac{SO-FSCL} approaches that without resource constraints.

The degradation in the latency gain of \ac{SO-FSCL} at low parallelism is due to a relatively fixed latency difference compared to \ac{SO-SCL}. This difference primarily stems from the internal \ac{LLR} updates under special nodes that are omitted by \ac{SO-FSCL}. Although the $2^{\psi}$ elements in the \ac{LLR} vector $\bm{\alpha}_{\psi,\phi}$ at the $\psi$-th stage can be updated in parallel, special nodes are typically located at low stages, causing the latency of \ac{LLR} updates beneath these nodes remains relatively insensitive to changes in parallelism. For instance, for the $(512, 256)$ polar code, when $N_{\rm par}$ decreases from $32L$ to $2L$, the time steps required by \ac{LLR} updates increase from 1059 to 2397 for \ac{SO-SCL} and from 122 to 1044 for \ac{SO-FSCL}, while the latency difference only increases from 947 to 1353. The proportional increase of this difference is much lower than that of the overall latency. As a result, as parallelism decreases, the proportion of latency reduction for \ac{SO-FSCL} diminishes significantly.

\begin{table}
    \caption{Required Number of Time Steps for Different Polar Decoders with Resource Constraints ($L=4$)}
    \centering
    \renewcommand{\arraystretch}{1.1}
    \begin{tabular}{@{~}lccccc@{~}}
        \toprule
        $(N,K)$ & $N_{\rm par}$ & SO-SCL & SO-FSCL & FastSOL & FSCL \\
        [1pt]\toprule
        & $2L$ & 1126 & 489 & 1131 & 414 \\
        (256,85) & $8L$ & 688 & 178 & 367 & 148 \\
        & $32L$ & 612 & 109 & 215 & 95 \\
        \midrule
        & $2L$ & 2727 & 1248 & 2779 & 1126 \\
        (512,256) & $8L$ & 1541 & 428 & 871 & 375 \\
        & $32L$ & 1323 & 240 & 471 & 216 \\
        \midrule
        & $2L$ & 5897 & 2700 & 6234 & 2458 \\
        (1024,512) & $8L$ & 3188 & 869 & 1871 & 772 \\
        & $32L$ & 2668 & 436 & 915 & 396 \\
        \bottomrule
	\end{tabular}
    \label{tab:ts_limit}
	\vspace*{-10pt}
\end{table}

\subsection{Decoding Complexity}
\begin{table*}
\centering
\begin{threeparttable}
    \caption{Decoding Complexity of Different Polar Decoders for Decoding (512, 256) 5G Polar Code with $L=4$\tnote{$\dagger$}}
    \renewcommand{\arraystretch}{1.1}
    \setlength{\tabcolsep}{5pt}
    \begin{tabular}{@{~}llcccccccccccccccc@{~}}
        \toprule
        & & \multirow{2}{*}[-3pt]{\begin{tabular}{@{}c@{}} ADD/\\SUB \end{tabular}} & \multirow{2}{*}[-3pt]{\begin{tabular}{@{}c@{}} MUL/\\DIV \end{tabular}} & \multirow{2}{*}[-3pt]{Bit Op.} & \multirow{2}{*}[-3pt]{\begin{tabular}{@{}c@{}} $\HD(\cdot)$/\\Sign Op. \end{tabular}} & \multirow{2}{*}[-3pt]{Compare} & \multirow{2}{*}[-3pt]{$\ln(1\s{+}e^x)$} & \multirow{2}{*}[-3pt]{$\ln(e^x\s{-}1)$} & \multicolumn{3}{c}{PM Sort\tnote{$\ddagger$} ~\& Sel.} & \multicolumn{5}{c}{LLR Sort\tnote{$\ddagger$}} \\
        \cmidrule(l){10-12} \cmidrule(l){13-18}
        & & & & & & & & & $2L$ & $4L$ & $8L$ & 2 & 4 & 8 & 16 & 32 & 64 \\
        [1pt]\toprule
        \multirow{5}{*}{Exact} & SO-SCL & 44374 & 0 & 9824 & 53164 & 9166 & 18055 & 0 & 254 & 0 & 0 & 0 & 0 & 0 & 0 & 0 & 0 \\
        & SO-FSCL & 26070 & 35 & 14633 & 29548 & 4914 & 12100 & 13 & 69 & 2 & 1 & 0 & 28 & 24 & 12 & 8 & 4 \\
        & FastSOL & 89104 & 0 & 9388 & 145180 & 21283 & 42496 & 0 & 62 & 0 & 0 & 64 & 48 & 24 & 0 & 8 & 0 \\
        & FSCL & 21228 & 0 & 14633 & 27476 & 3376 & 9584 & 0 & 69 & 2 & 1 & 0 & 28 & 24 & 12 & 8 & 4 \\
        & Pyndiah & 21382 & 0 & 14633 & 28040 & 3453 & 9636 & 0 & 69 & 2 & 1 & 0 & 28 & 24 & 12 & 8 & 4 \\
        \midrule
        \multirow{5}{*}{HWF} & SO-SCL & 37036 & 4973 & 9824 & 53164 & 45276 & -- & -- & 254 & 0 & 0 & 0 & 0 & 0 & 0 & 0 & 0 \\
        & SO-FSCL & 21476 & 3381 & 14633 & 29517 & 29001 & -- & -- & 69 & 2 & 1 & 0 & 28 & 24 & 12 & 8 & 4 \\
        & FastSOL & 65219 & 4081 & 9388 & 145180 & 106275 & -- & -- & 62 & 0 & 0 & 64 & 48 & 24 & 0 & 8 & 0 \\
        & FSCL & 18126 & 2933 & 14633 & 27476 & 22544 & -- & -- & 69 & 2 & 1 & 0 & 28 & 24 & 12 & 8 & 4 \\
        & Pyndiah & 18236 & 2937 & 14633 & 28040 & 22725 & -- & -- & 69 & 2 & 1 & 0 & 28 & 24 & 12 & 8 & 4 \\
        \bottomrule
    \end{tabular}
    \begin{tablenotes}
        \item[$\dagger$] The complexity is measured as the number of various operations required by the decoder, counted at $E_b/N_0=3$ dB.
        \item[$\ddagger$] The number of comparisons required by the sorting process is not counted in the ``Compare" column.
    \end{tablenotes}
    \label{tab:complx}
	\vspace*{-10pt}
\end{threeparttable}
\end{table*}

Since the \ac{SO} functionality of \ac{SO-FSCL} decoding is an add-on extension to \ac{FSCL} decoding, the additional decoding complexity of \ac{SO-FSCL} compared to \ac{FSCL} originates from extracting soft information (i.e., (\ref{eq:lambdaH}), (\ref{eq:lambdaL}), (\ref{eq:Lapp_apprL}), and (\ref{eq:lambdaHWF})) and the parallel \ac{SCL} decoder that decodes only frozen bits when $\min\{3, N_s/2\}<K_s<N_s$. To evaluate the complexity of \ac{SO-FSCL}, we present the number of various operations executed by different polar decoders for (512, 256) 5G polar code in Table \ref{tab:complx}. The list size is $L=4$, and the \ac{HWF} version employs approximation (\ref{eq:jaclog}). The list required for Pyndiah’s approach is provided by \ac{FSCL} decoding. It can be observed that node-based fast decoding not only significantly reduces decoding latency but also decreases computational costs for most operations. Moreover, the computation of \acp{SO} in \ac{SO-FSCL} decoding introduces certain computational costs relative to \ac{FSCL} decoding and Pyndiah's approach, predominantly for addition, multiplication, comparison, and nonlinear functions. For example, setting the exact version of the \ac{SO-SCL} decoder as the baseline, the \ac{SO-FSCL} decoder achieves 41.3\% reduction in additions, 46.4\% reduction in comparisons, and 33\% reduction in softplus computations. Meanwhile, the \ac{FSCL} decoder saves 52.2\% in additions, 63.2\% in comparisons, and 46.9\% in the softplus computations, while the complexity cost of Pyndiah's approach compared to \ac{FSCL} is negligible. Similar trends can also be observed for the \ac{HWF} versions. However, due to the backward \ac{SCAN} decoding process, the complexity of \ac{FastSOL} is several times higher than that of other decoders.

\section{Simulation Results}
In this section, we evaluate the \ac{SO} performance of the proposed \ac{SO-FSCL} decoder. Subsequently, we highlight its significant applications in two iterative decoding scenarios: \ac{MIMO} systems and \ac{GLDPC} codes.

\subsection{SO Performance}
\begin{figure}[!t]
    \centering
    \includegraphics[width=0.48\textwidth]{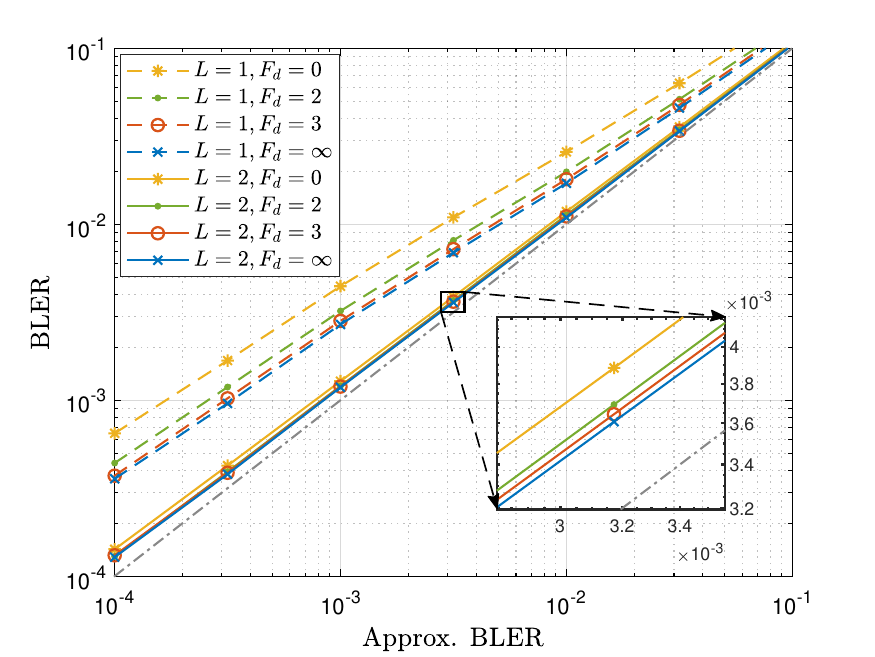}
    \caption{BLER vs. approximated BLER for $(512, 256)$ 5G polar code with different $F_d$ values over AWGN channel, evaluated at $E_b/N_0$ = 2.5 dB. The gray dash-dotted line represents the perfect approximation of BLER, while a greater deviation from it indicates a lower accuracy.}
    \label{fig:LER}
    \vspace{-10pt}
\end{figure}

We first discuss the choice of $F_d$ in \ac{FIC} nodes. As in \cite{Yuan2025SoSCL}, the dynamic frozen bits are generated through a convolutional transformation, essentially constituting the polarization-adjusted convolutional codes \cite{Arikan2019From}, by
\begin{equation}
    u[i] = \bigoplus_{j\in\mathcal{D}_i(\bm{g}^n)} u[j], i\in\mathcal{F},
\end{equation}
where $\mathcal{D}_i(\bm{g}^n) = \{j: j \in \Idx{i\s{-}n\s{+}1, i\s{-}1} \cap \mathcal{I}, \bm{g}^n[i\s{-}j\s{+}1] = 1\}$ and $\bm{g}^n$ is the generator
polynomial of constraint length $n-1$. Here, $\bm{g}^7 = (1,0,1,1,0,1,1)$ \cite{Arikan2019From,Yuan2025SoSCL} is adopted. Since setting only the first $F_d$ frozen bits in each FIC node as dynamic would alter the codebook, we refer to the method in \cite{Yuan2025SoSCL} to evaluate the impact of different $F_d$ on \acp{SO} while avoiding the influence of codebook changes. By using the approximation $P_\mathcal{T}^*(\bm{y}^N)$, the probability of a decision $\hat{\bm{u}}^N$ being correct is estimated as
\begin{equation}
    \Gamma^*(\hat{\bm{u}}^N, \bm{y}^N) = \frac{P_{\bm{U}^N|\bm{Y}^N} \left(\hat{\bm{u}}^N|\bm{y}^N\right)}{\sum_{\bm{u}^N\in\mathcal{V}} P_{\bm{U}^N|\bm{Y}^N} \left(\bm{u}^N|\bm{y}^N\right) + P_\mathcal{T}^*(\bm{y}^N)}.
\end{equation}
We collect blocks for which $1-\Gamma^*(\hat{\bm{u}}^N, \bm{y}^N)$ falls within the intervals $[10^{-j-0.1}, 10^{-j+0.1})$, $j=1,1.5,2,\cdots,4$, and compute both the \ac{BLER} and the expectation $E[1-\Gamma^*(\hat{\bm{u}}^N, \bm{y}^N)]$ for each interval. Fig. \ref{fig:LER} depicts the actual \ac{BLER} as a function of the approximated \ac{BLER} with different $F_d$ values, where $F_d=\infty$ implies that no restriction is imposed on the number of dynamic frozen bits. The results demonstrate that dynamic frozen bits contribute to a more accurate $P_\mathcal{T}^*(\bm{y}^N)$. For further details on how dynamic improves the approximation, readers may refer to Figs. 2 and 3 in \cite{Yuan2025SoSCL}. Moreover, Fig. \ref{fig:LER} indicates that larger $F_d$ values yield more accurate $P_\mathcal{T}^*(\bm{y}^N)$, while setting $F_d=3$ not only incurs negligible loss in \ac{SO} performance but also suffices to achieve compatibility with dynamic frozen bits through a look-up table.

Then, we compare the \ac{BER} performance of \ac{SO-SCL} and \ac{SO-FSCL} as \ac{APP} decoders, which estimate each bit according to $\hat{c}[i] = \arg\max_{a\in\{0,1\}} P_{C[i]|\bm{Y}^N}(a|\bm{y}^N)$, or equivalently, by performing hard decision on the \ac{APP} \ac{LLR} $\ell_{\text{APP}}[i]$. Table \ref{tab:ber} shows that both \ac{SO-FSCL} and \ac{SO-SCL} exhibit nearly identical \ac{BER} performance under both the exact version and the \ac{HWF} version. The estimation of \ac{HWF} version results in a slight degradation on \ac{BER}, while the performance difference caused by setting $F_d=3$ is negligible. Furthermore, without the modification from (\ref{eq:XiM}), the performance of \ac{SO-SCL} and \ac{SO-FSCL} experiences significant degradation, and this loss will be further exacerbated in iterative scenarios.

\begin{table*}
\centering
\begin{threeparttable}
    \caption{Comparison of BER Performance Between SO-SCL and SO-FSCL Decoders\tnote{$\dagger$}}
    \renewcommand{\arraystretch}{1.1}
    \setlength{\tabcolsep}{4pt}
    \begin{tabular}{@{~}cllccccccc@{~}}
        \toprule
        \multicolumn{3}{c}{$E_b/N_0$} & 1.5 dB & 1.75 dB & 2 dB & 2.25 dB & 2.5 dB & 2.75 dB & 3 dB \\
        \midrule
        & & Exact & $1.184\s{\times}10^{-2}$ & $5.334\s{\times}10^{-3}$ & $2.234\s{\times}10^{-3}$ & $8.799\s{\times}10^{-4}$ & $3.366\s{\times}10^{-4}$ & $1.394\s{\times}10^{-4}$ & $6.281\s{\times}10^{-5}$ \\
        & SO-SCL & Exact, w/o (\ref{eq:XiM}) & $1.330\s{\times}10^{-2}$ & $6.076\s{\times}10^{-3}$ & $2.559\s{\times}10^{-3}$ & $9.991\s{\times}10^{-4}$ & $3.791\s{\times}10^{-4}$ & $1.536\s{\times}10^{-4}$ & $6.694\s{\times}10^{-5}$ \\
        \multirow{2}{*}{(512, 256)} & & HWF & $1.222\s{\times}10^{-2}$ & $5.486\s{\times}10^{-3}$ & $2.270\s{\times}10^{-3}$ & $8.934\s{\times}10^{-4}$ & $3.331\s{\times}10^{-4}$ & $1.386\s{\times}10^{-4}$ & $6.395\s{\times}10^{-5}$\\ \cmidrule(l){2-10}
        $L=2$ & & Exact & $1.182\s{\times}10^{-2}$ & $5.336\s{\times}10^{-3}$ & $2.234\s{\times}10^{-3}$ & $8.803\s{\times}10^{-4}$ & $3.397\s{\times}10^{-4}$ & $1.390\s{\times}10^{-4}$ & $6.306\s{\times}10^{-5}$ \\
        & \multirow{2}{*}{SO-FSCL\hspace{-1cm}} & Exact, w/o (\ref{eq:XiM}) & $1.329\s{\times}10^{-2}$ & $6.087\s{\times}10^{-3}$ & $2.562\s{\times}10^{-3}$ & $1.000\s{\times}10^{-3}$ & $3.827\s{\times}10^{-4}$ & $1.532\s{\times}10^{-4}$ & $6.734\s{\times}10^{-5}$ \\
        & & HWF & $1.210\s{\times}10^{-2}$ & $5.462\s{\times}10^{-3}$ & $2.263\s{\times}10^{-3}$ & $8.965\s{\times}10^{-4}$ & $3.304\s{\times}10^{-4}$ & $1.387\s{\times}10^{-4}$ & $6.394\s{\times}10^{-5}$ \\
        & & HWF, $F_d=3$ & $1.175\s{\times}10^{-2}$ & $5.246\s{\times}10^{-3}$ & $2.177\s{\times}10^{-3}$ & $8.681\s{\times}10^{-4}$ & $3.297\s{\times}10^{-4}$ & $1.323\s{\times}10^{-4}$ & $6.434\s{\times}10^{-5}$ \\
        \bottomrule
        \toprule
        \multicolumn{3}{c}{$E_b/N_0$} & 1 dB & 1.25 dB & 1.5 dB & 1.75 dB & 2 dB & 2.25 dB & 2.5 dB \\
        \midrule
        & & Exact & $3.844\s{\times}10^{-2}$ & $1.738\s{\times}10^{-2}$ & $6.681\s{\times}10^{-3}$ & $2.169\s{\times}10^{-3}$ & $6.733\s{\times}10^{-4}$ & $1.927\s{\times}10^{-4}$ & $6.355\s{\times}10^{-5}$ \\
        & SO-SCL & Exact, w/o (\ref{eq:XiM}) & $4.636\s{\times}10^{-2}$ & $2.237\s{\times}10^{-2}$ & $9.052\s{\times}10^{-3}$ & $3.064\s{\times}10^{-3}$ & $9.738\s{\times}10^{-4}$ & $2.795\s{\times}10^{-4}$ & $8.802\s{\times}10^{-5}$ \\
        \multirow{2}{*}{(1024, 512)} & & HWF & $4.059\s{\times}10^{-2}$ & $1.848\s{\times}10^{-2}$ & $7.062\s{\times}10^{-3}$ & $2.286\s{\times}10^{-3}$ & $6.803\s{\times}10^{-4}$ & $1.988\s{\times}10^{-4}$ & $6.410\s{\times}10^{-5}$ \\ \cmidrule(l){2-10}
        $L=2$ & & Exact & $3.836\s{\times}10^{-2}$ & $1.738\s{\times}10^{-2}$ & $6.659\s{\times}10^{-3}$ & $2.173\s{\times}10^{-3}$ & $6.758\s{\times}10^{-4}$ & $1.923\s{\times}10^{-4}$ & $6.365\s{\times}10^{-5}$ \\
        & \multirow{2}{*}{SO-FSCL\hspace{-1cm}} & Exact, w/o (\ref{eq:XiM}) & $4.632\s{\times}10^{-2}$ & $2.236\s{\times}10^{-2}$ & $9.021\s{\times}10^{-3}$ & $3.067\s{\times}10^{-3}$ & $9.739\s{\times}10^{-4}$ & $2.795\s{\times}10^{-4}$ & $8.748\s{\times}10^{-5}$ \\
        & & HWF & $4.035\s{\times}10^{-2}$ & $1.833\s{\times}10^{-2}$ & $6.992\s{\times}10^{-3}$ & $2.277\s{\times}10^{-3}$ & $6.898\s{\times}10^{-4}$ & $1.981\s{\times}10^{-4}$ & $6.417\s{\times}10^{-5}$ \\
        & & HWF, $F_d=3$ & $3.969\s{\times}10^{-2}$ & $1.816\s{\times}10^{-2}$ & $6.884\s{\times}10^{-3}$ & $2.173\s{\times}10^{-3}$ & $6.917\s{\times}10^{-4}$ & $1.968\s{\times}10^{-4}$ & $6.249\s{\times}10^{-5}$ \\
        \bottomrule
    \end{tabular}
    \begin{tablenotes}
        \item[$\dagger$] In this table, $F_d=\infty$ unless otherwise specified.
    \end{tablenotes}
    \label{tab:ber}
	\vspace*{-10pt}
\end{threeparttable}
\end{table*}

\begin{figure}[!t]
    \centering
    \includegraphics[width=0.48\textwidth]{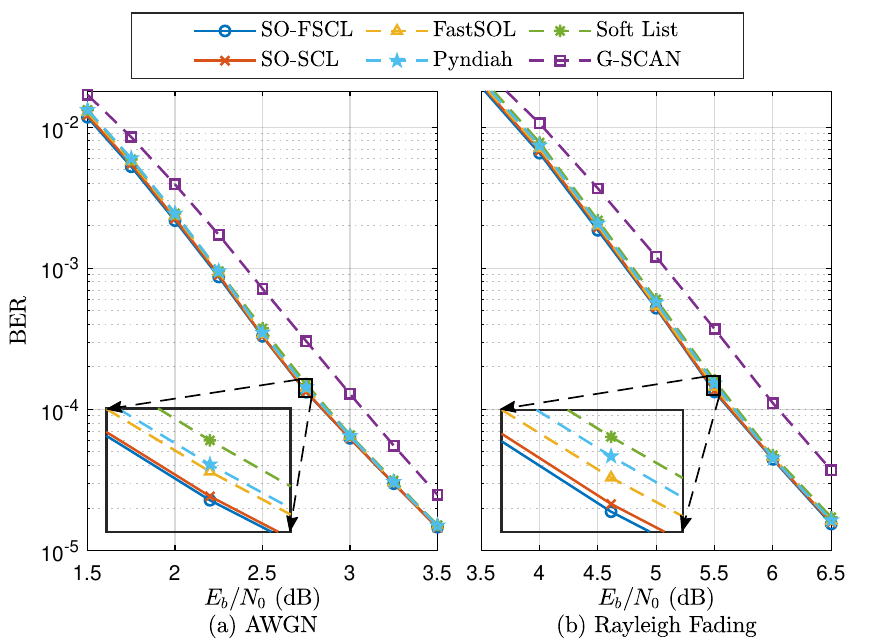}
    \caption{BER performance of various SO polar decoders for (512, 256) 5G polar code over AWGN and Rayleigh fading channels.}
    \label{fig:SO}
    \vspace{-10pt}
\end{figure}

The comparison of \ac{APP} decoding performance under different \ac{SO} polar decoders over both AWGN and Rayleigh fading channels is displayed in Fig. \ref{fig:SO}. The considered Soft List decoder \cite{Xiang2020Soft}, G-SCAN decoder \cite{Egilmez2022soft}, Pyndiah’s approach \cite{Pyndiah1998Near,Coskun2024Prec}, \ac{SO-SCL} decoder \cite{Yuan2025SoSCL}, and proposed \ac{SO-FSCL} decoder all adopt the \ac{HWF} version and set $L=2$, with $F_d = 3$ for \ac{SO-FSCL} and $F_d = \infty$ for the others. We determine the saturation value of Pyndiah’s approach according to \cite{Coskun2024Prec}. It is observed that \ac{SO-FSCL} decoder shows no performance loss compared with \ac{SO-SCL} decoder and outperforms other \ac{SO} decoders. It should be noted that although the \ac{BER} performance of the presented \ac{SO} decoders is very close, good \ac{BER} performance does not necessarily imply that the \acp{SO} are accurate, whereas accurate \acp{SO} will definitely lead to good \ac{BER} performance. This is because the \ac{APP} decoder only makes decisions on the signs of \acp{SO}. For example, the Soft List decoder uses the \ac{SCL} decoding list to correct the signs of \acp{SO}, ensuring that its performance is at least on par with that of \ac{SCL} \cite{Xiang2020Soft}. Moreover, if the sole objective is to obtain the estimated codewords, the mature \ac{SCL}/\ac{FSCL} decoder is undoubtedly a more suitable choice. The role of \acp{SO} and the impact of their accuracy will be more clearly demonstrated in the subsequent iterative decoding scenarios.

\subsection{Application 1: Iterative Detection and Decoding of MIMO Systems}
\ac{MIMO} is a revolutionary technology that multiplies the spectral efficiency of modern communication systems, and its channel capacity can be progressively approached through iterative detection and decoding \cite{Hochwald2003Ach}. To further illustrate the performance of proposed \ac{SO-FSCL} decoder, we apply various \ac{SO} decoders to $4\times4$ \ac{MIMO} iterative systems with \ac{QPSK} input. Each element of the channel matrix follows an independent and identically distributed Gaussian distribution with zero mean and unit variance. At the receiver, the channel state information is assumed to be perfectly known and a Max-Log \ac{MAP} detector \cite{robertson1995comparison} is used. 

\begin{figure*}[!t]
    \centering
    \includegraphics[width=\textwidth]{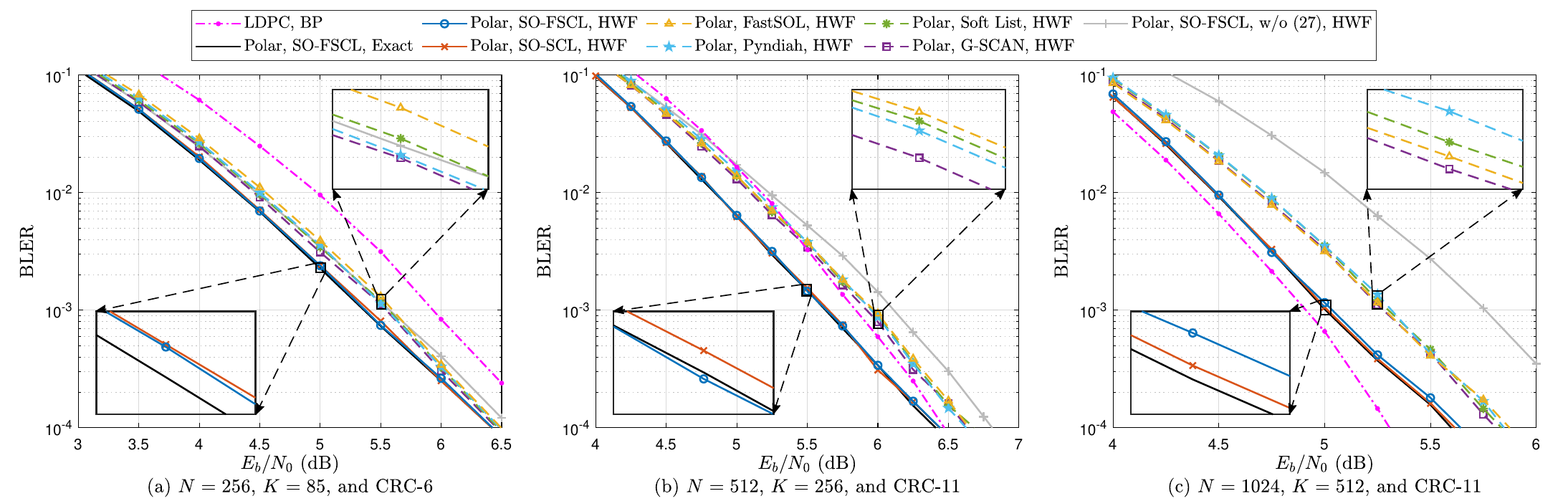}
    \caption{BLER performance of QPSK-input $4\times4$ MIMO iterative systems with various SO decoders for different 5G codes. The maximum number of iterative detection and decoding is set to $I=5$.}
    \label{fig:MIMO}
    \vspace{-10pt}
\end{figure*}

Fig. \ref{fig:MIMO} shows the \ac{BLER} performance of polar-coded and LDPC-coded \ac{MIMO} iterative systems with different \ac{SO} decoders. For 5G polar codes, the list size is set to $L=4$, while for 5G LDPC codes, the number of \ac{BP} decoding iterations is set to $I_{\rm BP}=12$ to achieve comparable complexity. The maximum number of outer iterations is set to $I = 5$. Analogously to Table \ref{tab:ber} and Fig. \ref{fig:SO}, the \ac{SO-FSCL} decoder exhibits almost identical performance to that of \ac{SO-SCL} decoder, with negligible performance loss in its \ac{HWF} version. Owing to more accurate \acp{SO}, the gap between \ac{SO-SCL}/\ac{SO-FSCL} and other \ac{SO} polar decoders has now widened in the \ac{MIMO} iterative systems. Moreover, the significant performance degradation of \ac{SO-SCL} without (\ref{eq:XiM}) demonstrates the necessity of the proposed modification. In line with the general consensus, the comparison between LDPC and polar codes confirms that polar-coded \ac{MIMO} systems exhibit superior performance over their LDPC counterparts as the code length decreases.

\begin{figure}[!t]
    \centering
    \includegraphics[width=0.48\textwidth]{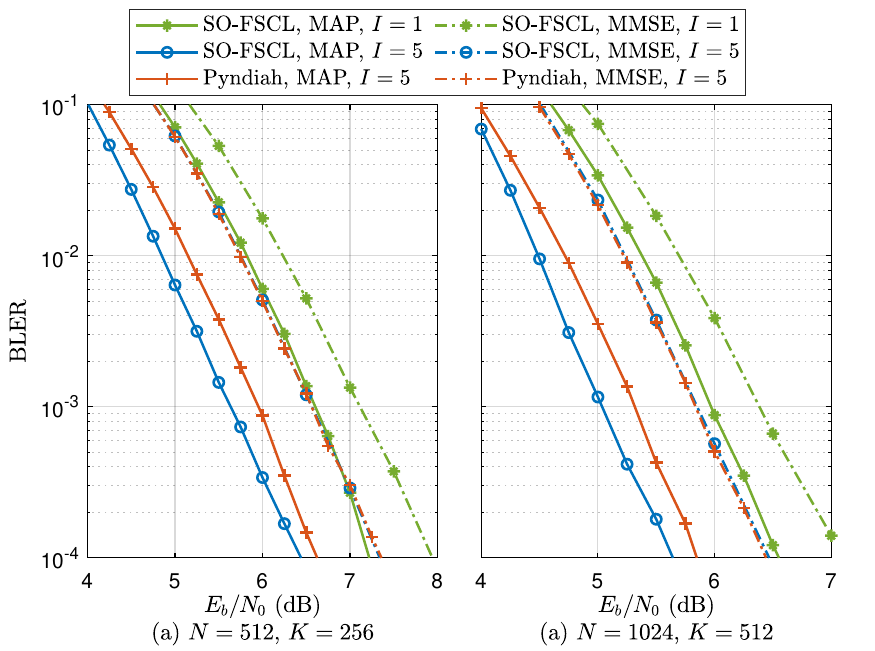}
    \caption{BLER performance of QPSK-input $4\times4$ MIMO iterative systems under MAP detection and MMSE detection for different 5G polar codes. Note that for the non-iterative case ($I=1$), Pyndiah's approach share the same performance with SO-FSCL, and thus is omitted from the figure.}
    \label{fig:MMSE}
    \vspace{-10pt}
\end{figure}

In Fig. \ref{fig:MMSE}, we further investigate the \ac{MIMO} iterative gain under \ac{MAP} detection and more practical \ac{MMSE} detection \cite{Studer2011ASIC}. The \ac{SO-FSCL} and Pyndiah decoders use the \ac{HWF} version with $L=4$. We observe that the iterative gain achieved by the optimal \ac{MAP} detector is more than twice that of the linear \ac{MMSE} detector. However, since the linear estimation of the \ac{MMSE} detector reduces the sensitivity of the iterative process to \ac{SO} accuracy, the gain of \ac{SO-FSCL} over other \ac{SO} polar decoders vanishes. This indicates that to fully leverage \acp{SO}, the design of \ac{MIMO} detectors should maintain sensitivity to the priori probabilities of constellation points while reducing complexity.

\subsection{Application 2: Iterative Decoding of GLDPC Codes}
By replacing the single parity check of LDPC codes with stronger component code constraints, \ac{GLDPC} codes have demonstrated larger minimum distances, enhanced iterative decoding performance, and lower error floors \cite{Liva2008Quasi}. With \ac{SOGRAND}, \cite{Yuan2025Soft} highlighted the important role of accurate \ac{SO} component decoders in the decoding performance of \ac{GLDPC} codes and product codes (a special class of \ac{GLDPC}). Subsequently, \ac{SO-SCL} was specifically applied to the decoding of \ac{GLDPC}/product codes based on polar component codes \cite{Yuan2025SoSCL}. In this subsection, we demonstrate that \ac{SO-FSCL} can also be applied to this scenario. 

\begin{figure}[!t]
    \centering
    \includegraphics[width=0.48\textwidth]{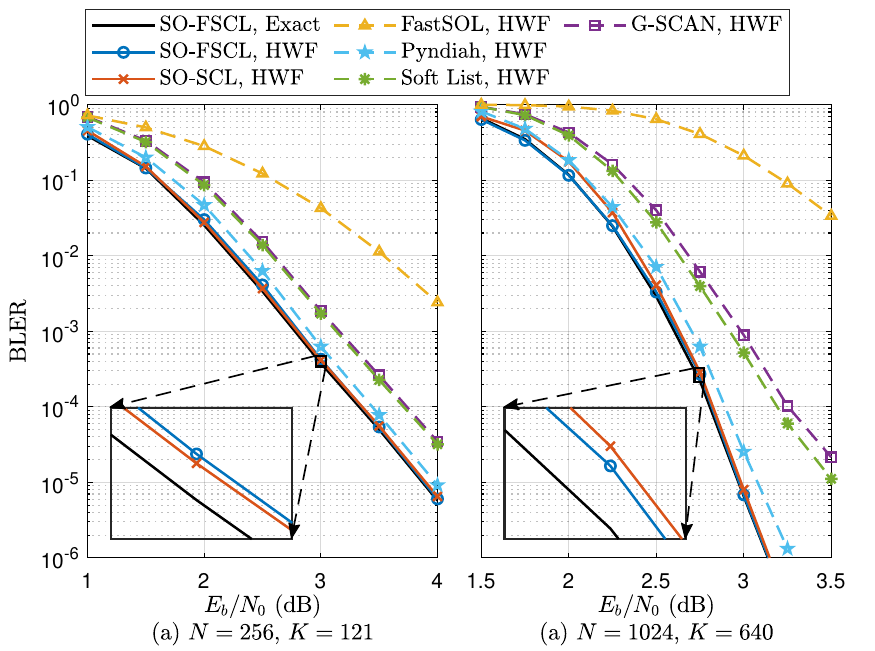}
    \caption{BLER performance of (256, 121) product code and (1024, 640) GLDPC code with different SO polar decoders over the AWGN channel.}
    \label{fig:GLDPC}
    \vspace{-10pt}
\end{figure}

We consider the (256, 121) product code based on (16, 11) polar code and the (1024, 640) \ac{GLDPC} code based on (32, 26) polar code. The product and \ac{GLDPC} codes based on \ac{eBCH} component codes with the same construction have been evaluated in \cite{Yuan2025Soft}, and the same \ac{GLDPC} code has also been evaluated in \cite{Yuan2025SoSCL}, where the \ac{GLDPC} code is constructed according to \cite{Michael2010From}. \mbox{Fig. \ref{fig:GLDPC}} presents the iterative decoding performance of these two codes with different \ac{SO} polar decoders over the AWGN channel. For the compared \ac{SO} polar decoders, the list size is $L = 4$ and the maximum number of iterations for the outer \ac{BP} algorithm is $I_{\rm BP} = 20$. During the iterative process, the scaling factors for extrinsic information passing of various \ac{SO} polar decoders are listed in Table \ref{tab:factor}, which are optimized in steps of 0.05. The simulation results show that under stricter requirements for \ac{SO} accuracy, BP/SCAN-based \ac{SO} decoders experienced more severe performance degradation when decoding \ac{GLDPC} codes. In contrast, \ac{SO-FSCL} (both exact and \ac{HWF} versions) maintains performance almost identical to \ac{SO-SCL}, while outperforming other decoders. 

\begin{table}[!t]
    \caption{Scaling Factors for Extrinsic Information Passing of Various SO Polar Decoders}
    \centering
    \scriptsize
    \renewcommand{\arraystretch}{1.1}
    \setlength{\tabcolsep}{3pt}
    \begin{tabular}{@{}cccccccc@{}}
    \toprule
    Code & Opt. @ & SO-FSCL & SO-SCL & FastSOL & Pyndiah & Soft List & G-SCAN \\ \midrule
    (256,121)  & 3.5 dB & 0.4 & 0.45 & 0.7 & 0.4 & 0.4 & 0.4 \\
    (1024,640) & 3 dB   & 0.55 & 0.55 & 0.7 & 0.55 & 0.55 & 0.55 \\ \bottomrule
    \end{tabular}
    \label{tab:factor}
	\vspace*{-10pt}
\end{table}

\begin{figure}[!t]
    \centering
    \includegraphics[width=0.48\textwidth]{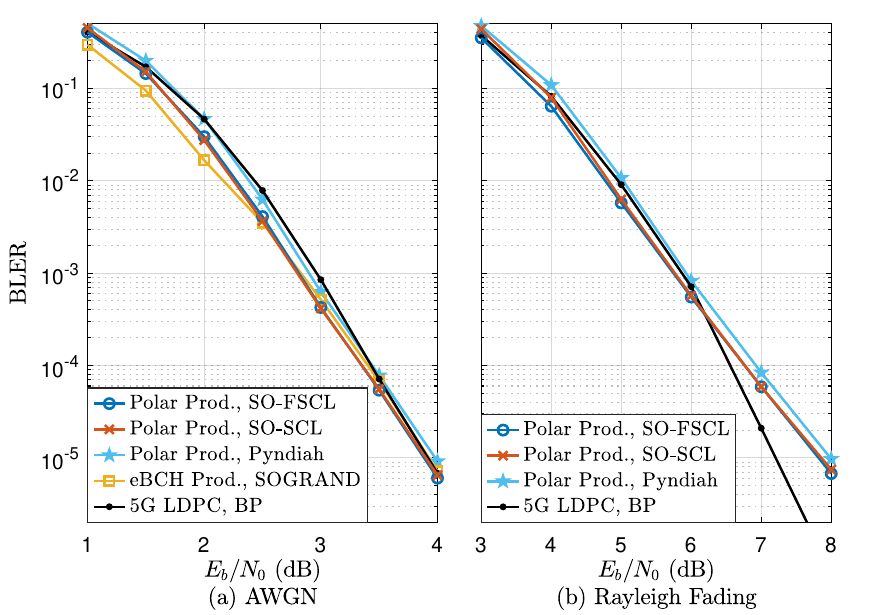}
    \caption{Performance comparison of polar product codes and other codes over AWGN and Rayleigh fading channels for $N=256$ and $K=121$.}
    \label{fig:GLDPC1}
    \vspace{-10pt}
\end{figure}

\begin{figure}[!t]
    \centering
    \includegraphics[width=0.48\textwidth]{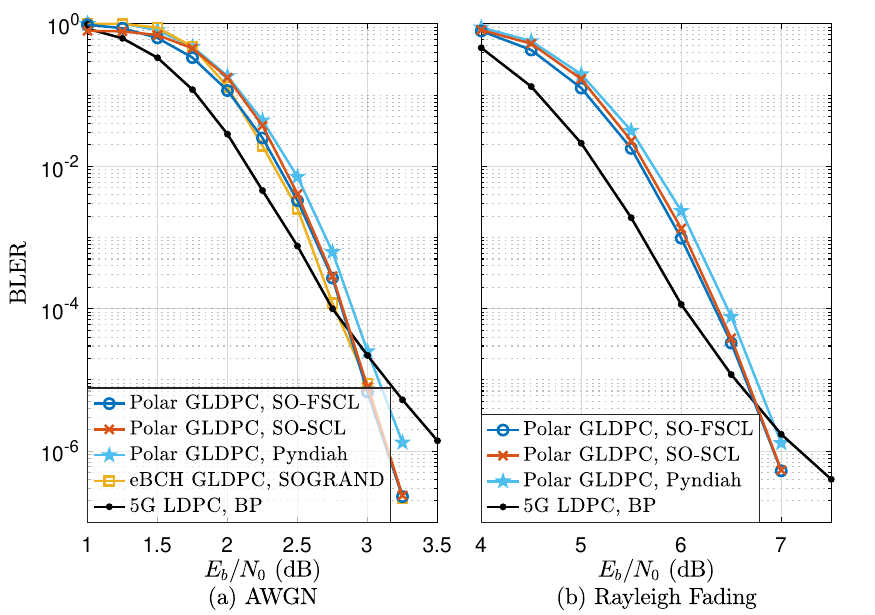}
    \caption{Performance comparison of polar GLDPC codes and other codes over AWGN and Rayleigh fading channels for $N=1024$ and $K=640$.}
    \label{fig:GLDPC2}
    \vspace{-10pt}
\end{figure}

In Figs. \ref{fig:GLDPC1} and \ref{fig:GLDPC2}, we further present a comparison of polar product/\ac{GLDPC} codes against \ac{eBCH} product/\ac{GLDPC} codes and 5G LDPC codes. The \ac{SO} polar decoders apply \ac{HWF} version with $L=4$, the curves for the \ac{eBCH} product/\ac{GLDPC} codes are sourced from \cite{Yuan2025Soft}, and the maximum number of BP iterations for 5G LDPC codes is $I_{BP} = 50$. Over the AWGN channel, the performance of polar product/\ac{GLDPC} codes is comparable to that of \ac{eBCH} product/\ac{GLDPC} codes and outperforms 5G LDPC codes in certain $E_b/N_0$ ranges. However, in fading channels, the gain over 5G LDPC codes experiences some degradation. Nevertheless, it must be emphasized that the performance of polar \ac{GLDPC} codes depends not only on accurate \ac{SO} component decoders but also on code construction. These remain ongoing research topics, and some preliminary results have been presented in \cite{Shi2025GLDPC}.

\section{Conclusion}
This paper demonstrates that \ac{SO-SCL} decoding is compatible with existing latency reduction techniques for \ac{SCL} decoding. By identifying \ac{FIC} nodes, the proposed \ac{SO-FSCL} decoding can provide additional bit-wise \acp{SO} based on \ac{FSCL} with only minor increases in latency and complexity, while significantly reducing both latency and complexity compared to \ac{SO-SCL}. Despite being faster and with lower complexity, \ac{SO-FSCL} demonstrates nearly identical performance to \ac{SO-SCL}, whether applied in \ac{APP} decoding, \ac{MIMO} iterative systems, or \ac{GLDPC} decoding, and consistently outperforms other state-of-the-art \ac{SO} polar decoders.

\bibliographystyle{IEEEtran}
\bibliography{contents/refer}


\begin{IEEEbiography}[{\includegraphics[width=1in,height=1.3in,clip]{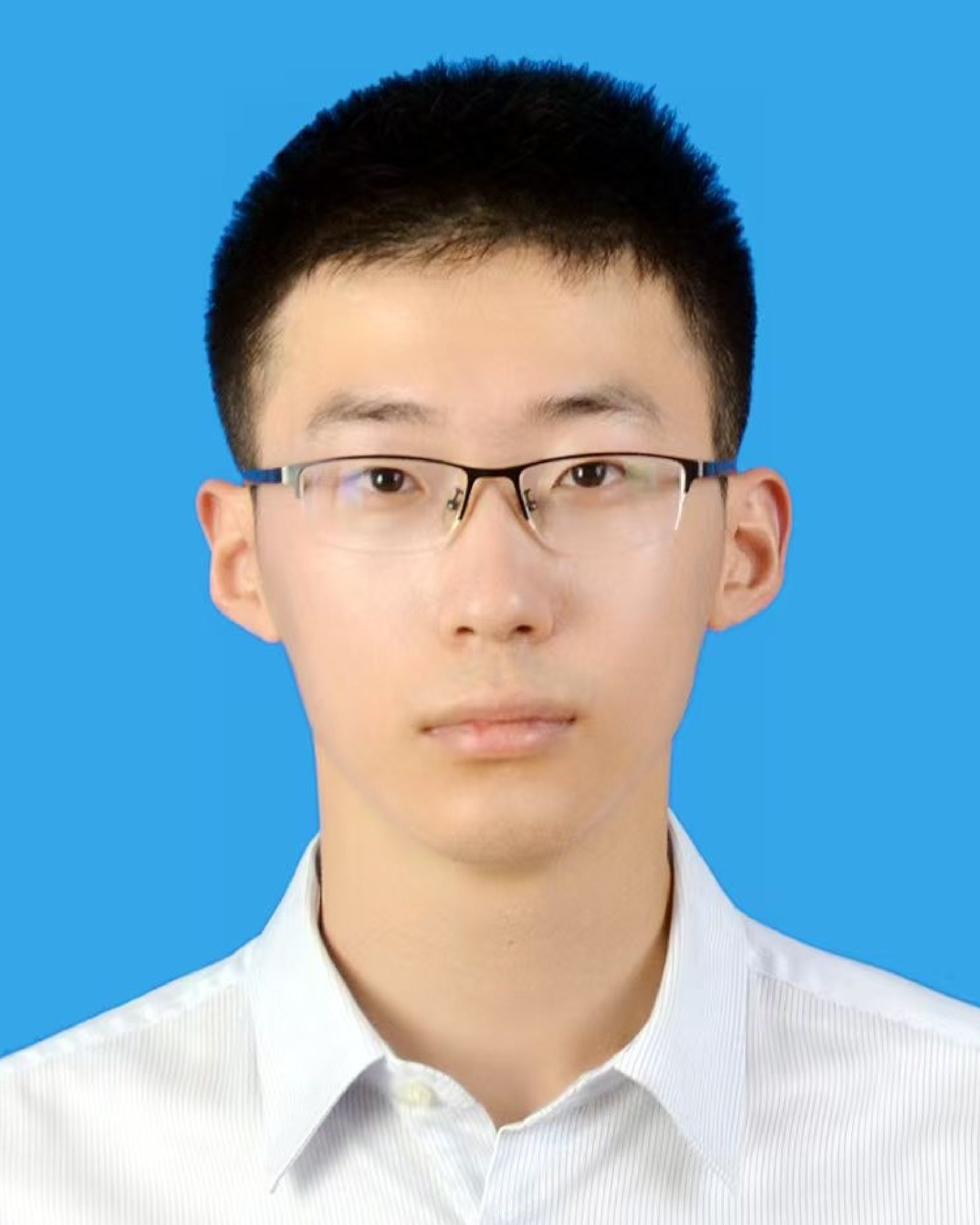}}]
{Li Shen}(Graduate Student Member, IEEE) received the B.S. degree in information engineering and the M.S. degree in electronic information from Shanghai Jiao Tong University, China, in 2021 and 2024, respectively. He is currently pursuing a Ph.D. degree in information and communication engineering with the School of Integrated Circuits (School of Information Science and Electronic Engineering), Shanghai Jiao Tong University. His research interests are channel coding, coded modulation, and physical layer security.
\end{IEEEbiography}

\begin{IEEEbiography}[{\includegraphics[width=1in,height=1.3in,clip]{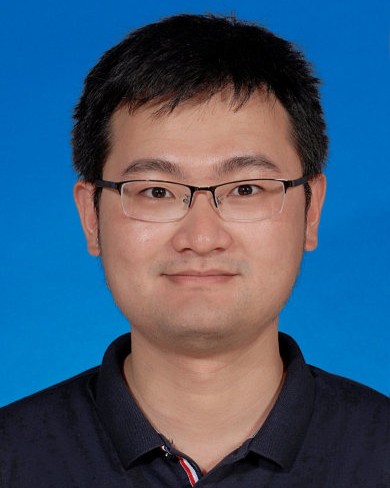}}]
{Yongpeng Wu}(Senior Member, IEEE) received the B.S. degree in telecommunication engineering from Wuhan University, Wuhan, China, in July 2007, and the Ph.D. degree in communication and signal processing from the National Mobile Communications Research Laboratory, Southeast University, Nanjing, China, in November 2013.

He is currently a Professor with the School of Integrated Circuits (School of Information Science and Electronic Engineering), Shanghai Jiao Tong University, China. Previously, he was a Senior Research Fellow with the Institute for Communications Engineering, Technical University of Munich, Germany, and the Humboldt Research Fellow, and a Senior Research Fellow with the Institute for Digital Communications, University Erlangen-N\"{u}rnberg, Germany. During his doctoral studies, he conducted cooperative research with the Department of Electrical Engineering, Missouri University of Science and Technology, USA. His research interests include massive MIMO/MIMO systems, massive machine type communications, physical layer security, and signal processing for wireless communications.

Dr. Wu was received the IEEE Student Travel Grants for IEEE International Conference on Communications (ICC) 2010, the Alexander von Humboldt Fellowship in 2014, the Travel Grants for IEEE Communication Theory Workshop 2016, the Excellent Doctoral Thesis Awards of China Communications Society 2016, the Exemplary Editor Award of IEEE Communication Letters 2017, the Young Elite Scientist Sponsorship Program by CAST 2017, and the Excellent Youth Science Fund Project of National Natural Science Foundation of China 2021. He has been the Symposium Chair of various conferences, including Globecom, ICC, VTC, and PIMRC. He has been the Lead Guest Editor of IEEE JOURNAL OF SELECTED TOPICS IN SIGNAL PROCESSING, IEEE JOURNAL ON SELECTED AREAS IN COMMUNICATIONS, and IEEE WIRELESS COMMUNICATIONS. He was an Editor of IEEE TRANSACTIONS ON COMMUNICATIONS and IEEE COMMUNICATIONS LETTERS. He is an Editor of IEEE TRANSACTIONS ON INFORMATION THEORY and IEEE WIRELESS COMMUNICATIONS.
\end{IEEEbiography}

\begin{IEEEbiography}[{\includegraphics[width=1in,height=1.3in,clip]{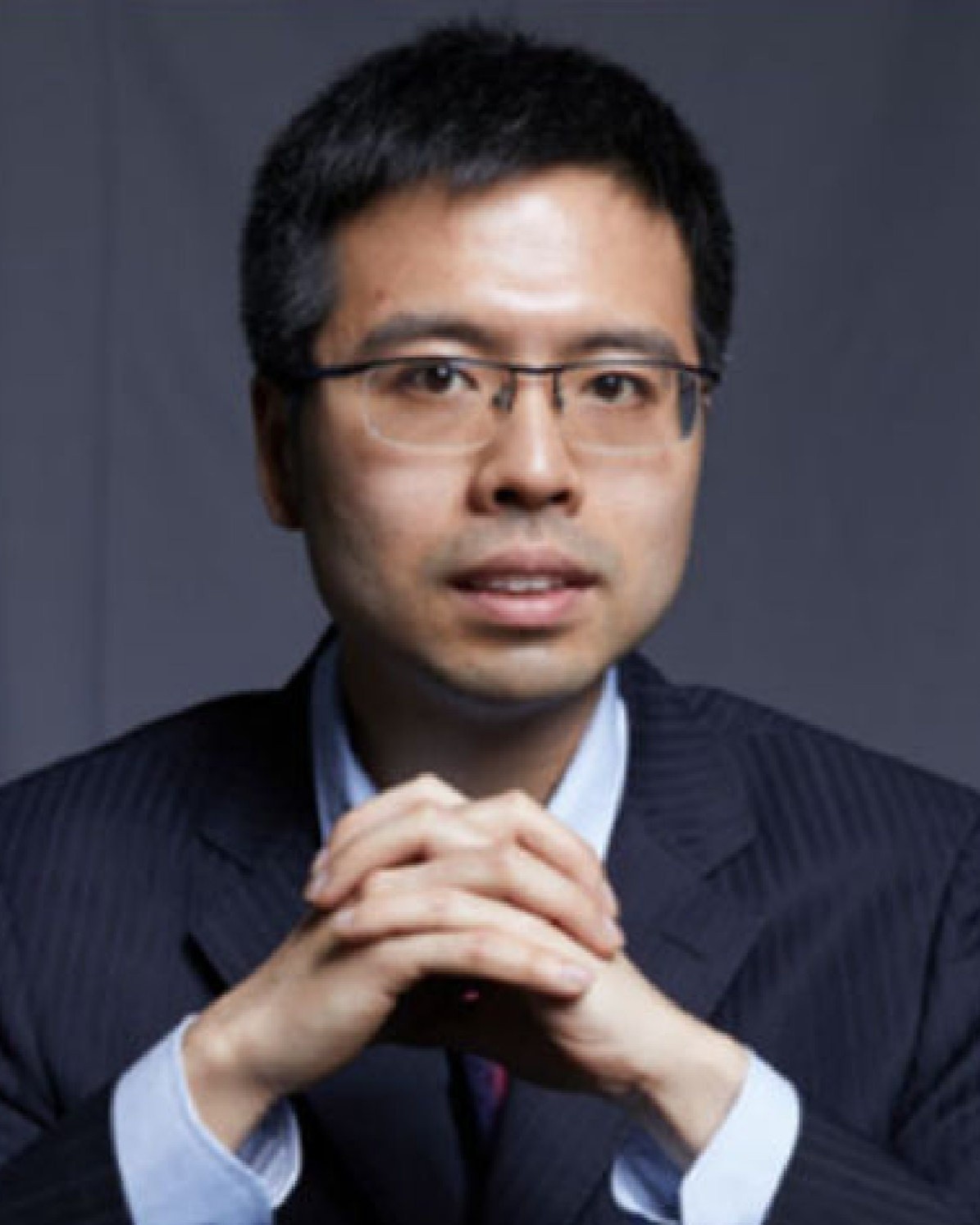}}]
{Zhen Gao}(Senior Member, IEEE) received the B.S. degree in information engineering from Beijing Institute of Technology, Beijing, China, in 2011, and the Ph.D. degree in communication and signal processing from Tsinghua National Laboratory for Information Science and Technology, Department of Electronic Engineering, Tsinghua University, China, in 2016. He is currently a Professor with Beijing Institute of Technology. His research interests include wireless communications, with a focus on multi-carrier modulations, multiple antenna systems, and sparse signal processing. He was a recipient of the IEEE Broadcast Technology Society 2016 Scott Helt Memorial Award (Best Paper), the Exemplary Reviewer of IEEE COMMUNICATION LETTERS in 2016, IET Electronics Letters Premium Award (Best Paper) 2016, and the Young Elite Scientists Sponsorship Program (2018–2021) from China Association for Science and Technology.
\end{IEEEbiography}

\begin{IEEEbiography}[{\includegraphics[width=1in,height=1.3in,clip]{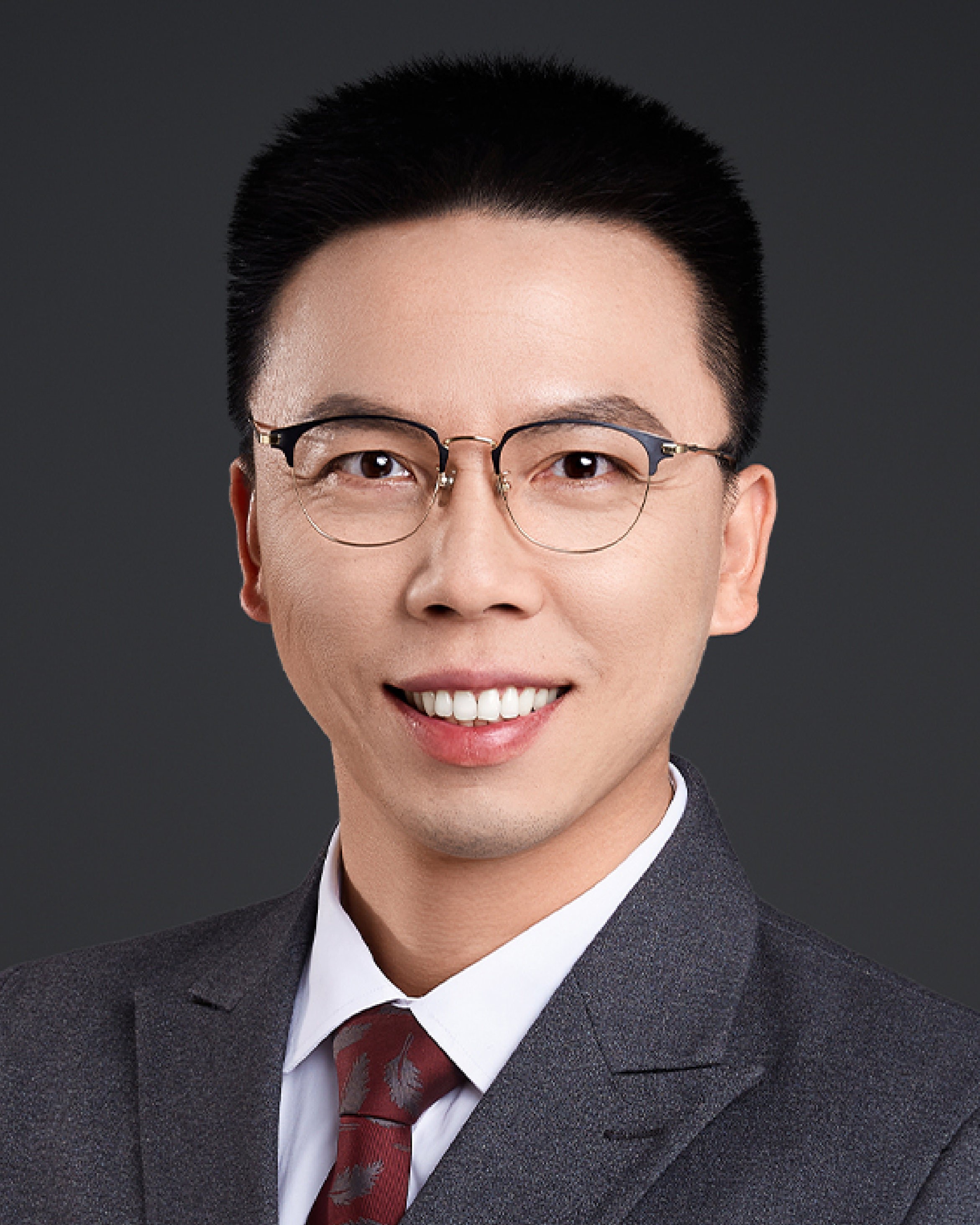}}]
{Yin Xu}(Senior Member, IEEE) received the B.Sc. degree in information science and engineering from Southeast University, China, in 2009, and the M.Sc. and Ph.D. degrees in communication and information systems from Shanghai Jiao Tong University (SJTU), China, in 2011 and 2015, respectively. In 2011, he was a Visiting Scholar at McGill University and INRS. He is currently an Associate Professor and a Ph.D. Supervisor with SJTU. His contributions to core physical layer technologies have been incorporated as key components into multiple globally reputable communication standards, including 3GPP 5G, ATSC 3.0, and SparkLink. His research interests include key technologies, such as channel coding and modulation, new waveforms, broadcast/multicast, and AI-assisted communications, along with the development of prototype systems for diverse communication systems, including 5G, broadcast, and short-range communications. He was a recipient of the 6G Star Youth Scientist Award. He also serves as the co-chair, the session chair, and a keynote speaker at various major IEEE international conferences. Notably, he is actively involved in communication system standardization, serving as the chief 3GPP Standard delegate representing SJTU and holding the vice-chair position of Implementation Team 5 under the American Advanced Television Systems Committee (ATSC). He serves as an Associate Editor for IEEE TRANSACTIONS ON BROADCASTING.
\end{IEEEbiography}

\begin{IEEEbiography}[{\includegraphics[width=1in,height=1.3in,clip]{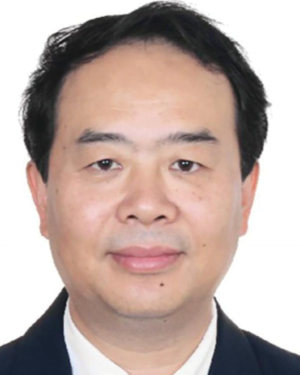}}]
{Xiaohu You}(Fellow, IEEE) received the master’s and Ph.D. degrees in electrical engineering from Southeast University, Nanjing, China, in 1985 and 1988, respectively. Since 1990, he has been with the National Mobile Communications Research Laboratory, Southeast University, where he holds the rank of the Director and a Professor. From 1999 to 2002, he was the Principal Expert of the C3G Project, responsible for organizing China’s 3G Mobile Communications Research and Development Activities. From 2001 to 2006, he was the Principal Expert of China National 863 Beyond 3G FuTURE Project. Since 2013, he has been the Principal Investigator of China National 863 5G Project. He has contributed over 200 IEEE journal articles and two books in the areas of adaptive signal processing and neural networks and their applications to communication systems, and made important contributions to the development of China’s 3G, 4G, and 5G mobile communications. His research interests include mobile communication systems and signal processing and its applications. He was selected as an IEEE Fellow for his contributions to the development of mobile communications in China in 2011. He was a recipient of the National First Class Invention Prize in 2011. He has served as the General Chair of IEEE WCNC 2013, IEEE VTC 2016 Spring, and IEEE ICC 2019. He is the Secretary General of the FuTURE Forum and the Vice Chair of China IMT-2020 Promotion Group and China National Mega Project on New Generation Mobile Network. He won the Chan Kah Kee Science Award in 2014 and the IET Achievement Medal in 2021. He was a Member of Chinese Academy of Sciences.
\end{IEEEbiography}

\begin{IEEEbiography}[{\includegraphics[width=1in,height=1.3in,clip]{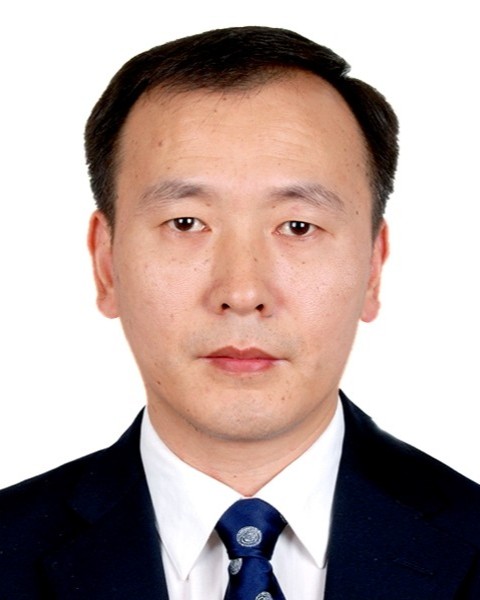}}]
{Xiqi Gao}(Fellow, IEEE) received the Ph.D. degree in electrical engineering from Southeast University, Nanjing, China, in 1997. He joined the Department of Radio Engineering, Southeast University, in April 1992. Since May 2001, he has been a professor of information systems and communications. From September 1999 to August 2000, he was a visiting scholar at Massachusetts Institute of Technology, Cambridge, and Boston University, Boston, MA. From August 2007 to July 2008, he visited the Darmstadt University of Technology, Darmstadt, Germany, as a Humboldt scholar. His current research interests include broadband multicarrier communications, massive MIMO wireless communications, satellite communications, optical wireless communications, information theory and signal processing for wireless communications. From 2007 to 2012, he served as an Editor for the IEEE Transactions on Wireless Communications. From 2009 to 2013, he served as an Associate Editor for the IEEE Transactions on Signal Processing. From 2015 to 2017, he served as an Editor for the IEEE Transactions on Communications.

Dr. Gao received the Science and Technology Awards of the State Education Ministry of China in 1998, 2006, 2009 and 2025, the National Technological Invention Award of China in 2011, the Science and Technology Award of Jiangsu Province of China in 2014, and the 2011 IEEE Communications Society Stephen O. Rice Prize Paper Award in the field of communications theory.
\end{IEEEbiography}

\begin{IEEEbiography}[{\includegraphics[width=1in,height=1.3in,clip]{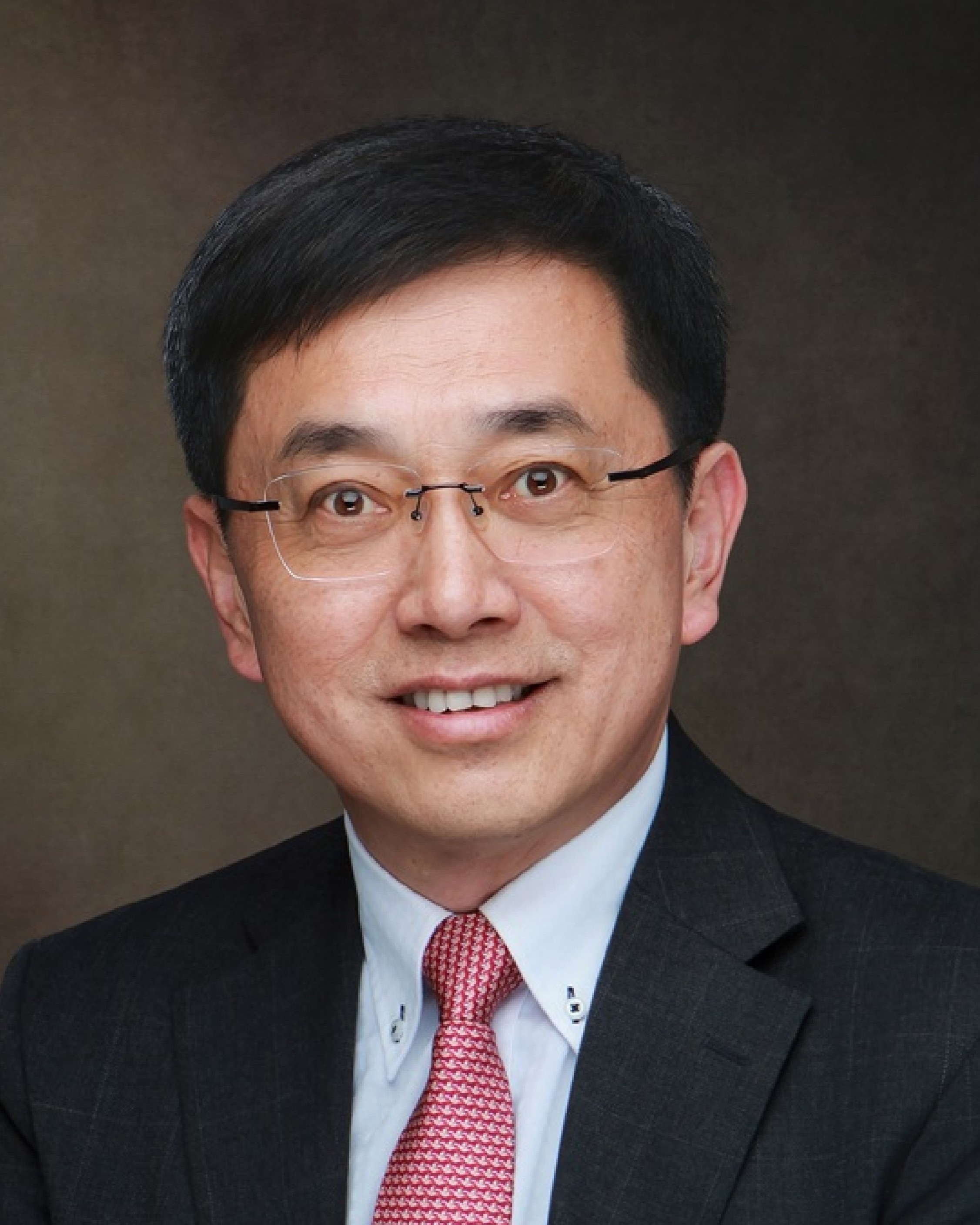}}]
{Wenjun Zhang}(Fellow, IEEE) received the B.S., M.S., and Ph.D. degrees in electronic engineering from Shanghai Jiao Tong University, Shanghai, China, in 1984, 1987, and 1989, respectively. After three years’ working as an Engineer with Philips, Nuremberg, Germany, he went back to his Alma Mater in 1993 and became a Full Professor in electronic engineering in 1995. He is one of the main contributors of the Chinese DTTB Standard (DTMB) issued in 2006. He is currently the Chief Scientist of Chinese Digital TV Engineering Research Centre (NERC-DTV), an industry/government consortium in DTV technology research and standardization, and the Director of the Cooperative Media Network Innovation Centre (CMIC), and an Excellence Research Cluster of media industry and academia. He is also an Academician of Chinese Academy of Engineering. He holds 240 patents and has published more than 170 papers in international journals and conferences. His main research interests include intelligent video processing/coding and transmission, multimedia semantic analysis, and broadcast/broadband network convergence.
\end{IEEEbiography}

\end{document}